\documentclass{aa}

\newcommand{\cd}{$\mathrm{d}^{-1}$}

\usepackage{color}
\usepackage{graphicx}
\usepackage{txfonts}
\usepackage{booktabs} 
\usepackage{isotope}

\begin{document} 

   \title{Tidally perturbed pulsations in the pre-main sequence $\delta$ Scuti binary RS Cha}

   \author{T. Steindl\inst{1}, K. Zwintz
          \inst{1}, D. M. Bowman \inst{2}}

   \institute{\inst{1}Institut f\"ur Astro- und Teilchenphysik, Universit\"at Innsbruck, Technikerstra{\ss}e 25, A-6020 Innsbruck, Austria\\
              \email{thomas.steindl@uibk.ac.at} \\
   \inst{2}Institute of Astronomy, KU Leuven, Celestijnenlaan 200D, B-3001 Leuven, Belgium\\
             }

   \date{Received 03.08.2020; accepted 16.11.2020}

 
  \abstract
   {Stellar components in binaries are subject to tidal forces which influence asteroseismic properties. Tidally pertubed pulsations have been reported for different objects but none of these are in their pre-main sequence phase of evolution. This makes RS Cha, consisting of two $\delta$ Scuti stars and with pulsational characteristics influenced by tidal effects , the first such object observed.}
   {We aim to investigate the pulsational properties of the eclipsing binary RS Cha in terms of the theory of tidally perturbed pulsations.}
   {Based on photometric time series obtained from the TESS satellite, we performed binary modelling using PHOEBE to interpret the binary light curve and to allow the investigation of the pulsations of both components in RS Cha. We modelled the detrended light curve with the superposition of linear modes. The frequencies were then interpreted as self excited modes perturbed by tidal forces.}
   {We find evidence for tidally perturbed modes, which enables the identification of pulsation modes. RS Cha mainly exhibits dipole modes, while one prominent $l=2$ or $l=3$ mode is also inferred. The latter verifies previous results from spectroscopic time series. }
   {This work shows that RS Cha is an ideal candidate to test the theory of tidally perturbed pulsations within the framework of asteroseismic modelling. The identification of multiple pulsation modes using this theory is unprecedented and will be a keystone in the future of pre-main sequence asteroseismology. However, amplitude modulation caused by the changing light ratio during the orbital phase in an eclipsing binary also plays a significant role, which can complicate mode identification.
   }


   \maketitle
%

\section{Introduction}
Asteroseismology -- the study of stellar pulsations -- is an important tool in studying stellar structure and evolution \citep{aerts2010}. The capabilities of asteroseismic tools are inherently dependent on the type of pulsators and therefore the evolutionary stage of the star. Stars with convective envelopes (e.g. red giants) are subject to the stochastic driving of pressure (p) modes which leads to solar-like oscillations that follow simple scaling relations, allowing for a robust determination of the star's mass and radius \citep[e.g.][]{chaplin2013, garcia2019, aerts2020}. Sophisticated methods of analysis provide better estimates that are important for the classifications of exoplanets \citep[e.g.][]{Lundkvist2018, Adibekyan2018}. For solar-like pulsators, the presence of acoustic glitches in frequency spacings allows the derivation of helium abundances even if stars do not display helium spectral lines \citep{verma2014, verma2017, verma2019a, verma2019b, aerts2020}. On the other hand, for more massive main sequence stars with convective cores, the techniques of asteroseismology rely on the exploitation of gravity (g) modes \citep[e.g.][]{VanReeth2016, VanReeth2018, Christophe2018, Mombarg2019, Bowman2020}.

Such detailed analyses are confined to stars on the main sequence or later evolutionary stages. High-quality observations are much less frequent among pre-main sequence stars owing to their fast evolution towards the main sequence and their sky position. Pre-main sequence stars are mainly located in crowded star-forming regions along the ecliptic, a region of the sky which the main Kepler mission \citep{koch2010} pointed away from \citep{zwintz2017}. Moreover, pre-main sequence stars are often subject to periodic or aperiodic variability caused by various types of activity such as from disks or magnetic fields, which further complicates carrying out a detailed analysis \citep{zwintz2019}. The pre-main sequence evolution is a vital building block of a star's life  connecting star formation to later evolutionary stages, starting from the zero-age main sequence. It therefore plays a major role concerning, for example, the transport of angular momentum \citep{zwintz2019}. 

A significant amount of known pre-main sequence pulsators are $\delta$ Scuti{-}type objects. They are located within the classical instability strip \citep{rodriguez2001} with effective temperatures in the range of $6300 \leq T_\mathrm{eff} \leq 9000$~K and luminosities in the range of $0.6 \leq \log (L/L_\odot) \leq 2.0$ \citep{buzasi2005}, corresponding to effective surface gravities in the range of $3.5 \leq \log(g) \leq 4.5$ at stellar masses in the range of approximately $1.5 \leq M/M_\odot \leq 3.5$ for pre-main sequence objects \citep{zwintz2014}. Typically, $\delta$ Scuti{-}type stars oscillate in radial and non-radial p modes and low-order g modes, exhibiting periods within the range of $18$~min to $8$~h \citep[e.g.][]{pamyatnykh2000, bowman2018, antoci2019}. \citet{zwintz2014} show the presence of a relationship between the pulsation properties and the evolutionary status for pre-main sequence stars. The recent work by \citet{bedding2020} indicates that a similar relationship also exists for young main sequence $\delta$ Scuti stars.

  \begin{table}
      \caption[]{Stellar and and orbital parameters for RS Cha.}
         \label{tab:rs_cha_params}
         
    \begin{tabular*}{\linewidth}{l@{\extracolsep{\fill}}lll}
            \hline
            \noalign{\smallskip}
            &  Primary &  Secondary &  Reference \\
            \noalign{\smallskip}
            \hline
            \noalign{\smallskip}
            $M/M_\odot$                             & $1.89(1)$     & $1.87(1)$         & A05     \\
                                                    & $1.823(12)$   & $1.764(12)$       & W13     \\
            $R/R_\odot$                             & $2.15(6)$     & $2.36(6)$         & A05     \\
            $T_\mathrm{eff}$ (K)                    & $7638(76)$    & $7228(72)$        & R00     \\
                                                    & $7926(150)$   & $7330(100)$       & F06     \\
            $\mathrm{log}(L/L_\odot)$               & $1.15(9)$     & $1.13(9)$         & A05     \\
            $\mathrm{log}(g)$ (cm $\mathrm{s}^{-2}$)   & $4.05(6)$     & $3.96(6)$         & A05    \\
            $v~\mathrm{sin}~i$ (km $\mathrm{s}^{-1}$)  & $64(6)$       & $70(6)$           & A05   \\
                                                    & $68(2)$      & $72(2)$       & W13       \\
            expected                                & $66(3)$       & $72(3)$   &    \\
            $P$ (days)                              & \multicolumn{2}{c}{$1.66988(2)$}  & W13     \\
            $i$ ($^\circ$)                          & \multicolumn{2}{c}{$83.4(3)$}       & CN80    \\
            $e~\mathrm{sin}~w$                          & \multicolumn{2}{c}{$-0.013(4)$}       & CN80    \\
            $[\mathrm{Fe}/\mathrm{H}]$              & \multicolumn{2}{c}{$0.017(1)$}    & A05     \\
            age (Myr)               & \multicolumn{2}{c}{$6^\mathrm{+2}_\mathrm{-1}$}   & LS04    \\ 
            \noalign{\smallskip}
            \hline
    \end{tabular*}    
    \tablefoot{
     For some parameters, multiple sources are given. \\
    References: A05: \citet{alecian2005}, W13: \citet{woollands2013}, R00: \citet{ribas2000}, F06: \citet{fremat2006} , CN80: \citet{clausen1980}, LS04: \citet{luhman2004}
    }
   \end{table}

Such intermediate mass stars often exist in binary systems, which leads to interactions having substantial effect on their structure and evolution (see e.g. \citealt{deamrco2017} for a review). These interactions moreover lead to deviations from sphericity and tidal effects, which in turn influence the seismic properties of the stars. For close binaries in an eccentric orbit, tidal effects can lead to the excitation of g-modes at multiples of the orbit frequency \citep{fuller2017}. This is typically observed in heartbeat stars \cite[e.g.][]{welsh2011} or eclipsing binaries \citep[e.g.][]{hambleton2013, maceroni2014}. \citet{handler2020} and \citet{kurtz2020} present the cases of HD 74423 and CO Cam, where tidal forces modified the orientation of the pulsation axis leading to tidal trapping of the oscillations \citep{Fuller2020}. 

Tidal effects further lead to a deformation of the pulsation cavity that results in a perturbation of self-excited pulsation modes \citep{polfliet1990, reyniers2003b, reyniers2003a}. In the observers' frame, this leads to multiplets spaced by (twice) the orbital frequency \citep{smeyers2005}. Such perturbed modes have been observed in a handful of stars \citep{lee2016, balona2018, bowman2019, Southworth2020, jerzykiewicz2020} and probably also in the eclipsing binary discussed by \citet{hambleton2013} as pointed out by \citet{balona2018}. 

In this work, we present an asteroseismic analysis of the pre-main sequence eclipsing binary system RS Cha, the first pre-main sequence objects showing evidence of tidal effects on the pulsation characteristics. This has been made possible by the stunning precision of light curves obtained by the \textit{Transiting Exoplanet Survey Satellite} \citep[TESS;][]{ricker2015}. In Sect. \ref{sec:system} we discuss the system RS Cha and the previous publications focusing on the orbital and stellar parameters as well as the pulsation characteristics. Section \ref{sec:observations} describes the observations that were used in the analysis. Section \ref{sec:binmod} describes the binary light curve modelling. We explain the frequency analysis and discuss evidence for tidal effects on the light curve of RS Cha in Sect. \ref{sec:pulsanal} and Sect. \ref{evidence}. We present our conclusions in Sect. \ref{sec:conclusion}. 
\begin{figure}
   \centering
   \includegraphics[width=\linewidth]{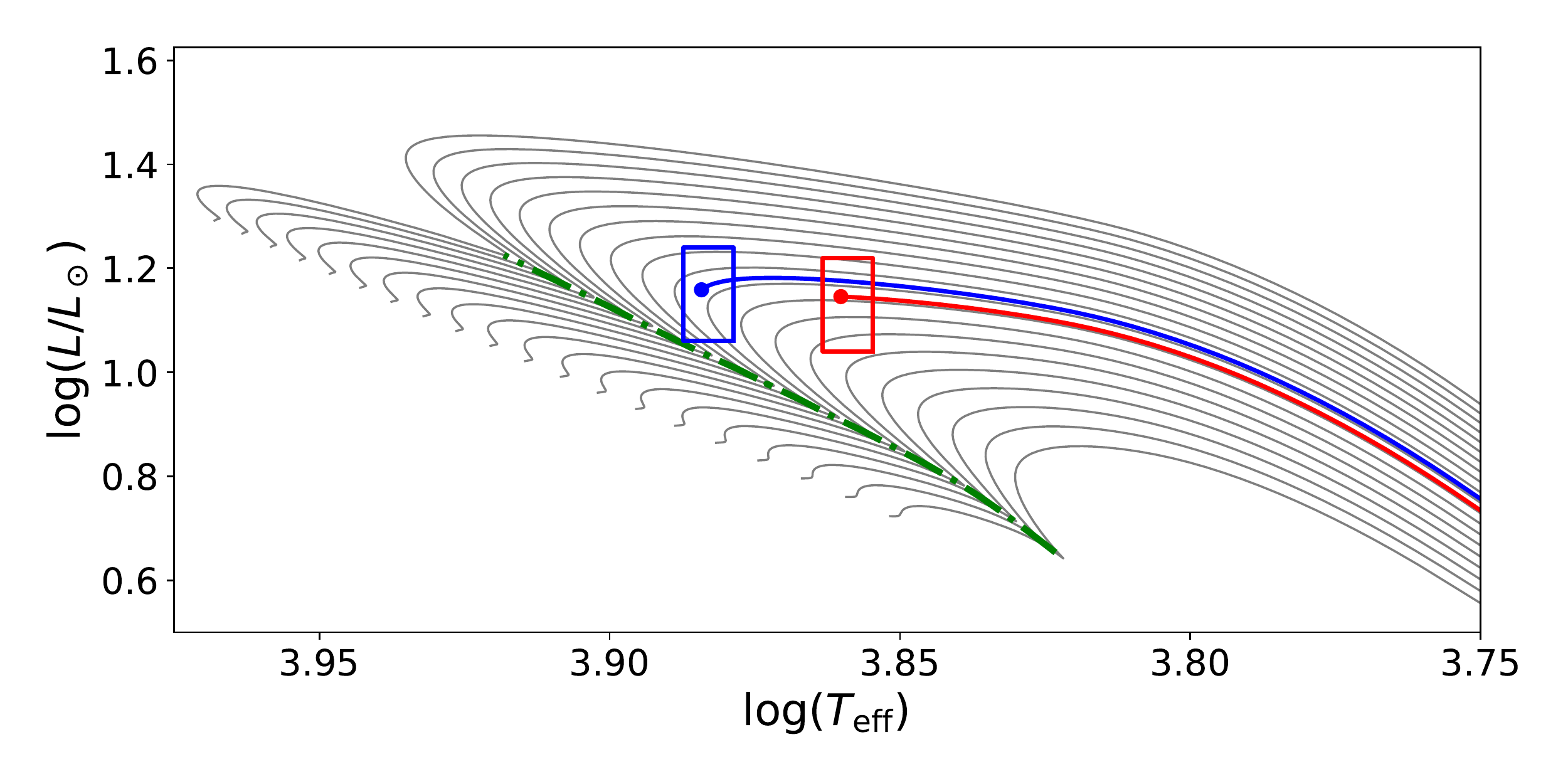}
      \caption{Evolutionary status of the components of RS Cha in a Hertzsprung{-}Russell diagram. The grey lines show the evolutionary tracks of $1.6$~$M_\odot$ to $2.0$~$M_\odot$ stars towards the end of their pre-main sequence phase. The blue and red box shows the observational constraints for the primary and secondary respectively. Also shown are evolutionary tracks corresponding to masses reported by \citet{alecian2005} that reproduce the observed radius at their final point (marked with a dot) for the primary (blue) and secondary (red). The green dash-dotted line marks the point of evolution where the central $\rm{C}^{12}$ mass fraction drops below $10^{-4}$.
              }
         \label{Fig:1:evolutionary_status}
\end{figure}

\begin{figure*}
   \centering
   \includegraphics[width=\linewidth]{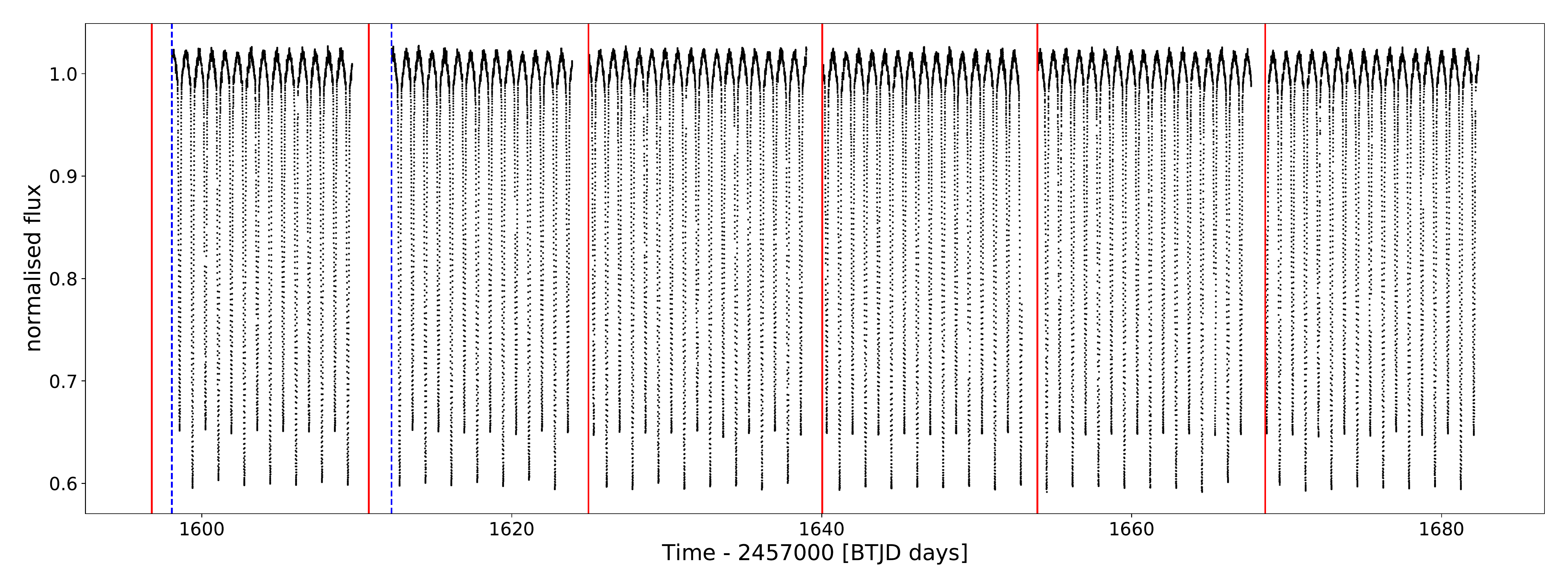}
      \caption{SAP FLUX light curve of RS Cha as black dots. Vertical red lines mark the start of any Orbits. Orbits 29 and 30 in sector 11 suffered from scattered light signals. The corresponding measurements have been removed by the TESS pipeline. This concerns the times between the start of the orbits and the blue dashed lines.
              }
         \label{Fig:1:Full_LC_p_gaps}
\end{figure*}
\section{The eclipsing binary system RS Cha}
\label{sec:system}
\subsection{Orbital and stellar parameters}
\citet{cousins1960} first reported RS Cha to be a variable star with visual magnitude in the range of $6.02$ to $6.05$ at a spectral type A5. Only four years later, \cite{strohmeier1964} was the first to discover the binary nature of RS Cha, denoting it as eclipsing binary of either Algol or $\beta$ Lyrae{-}type. Later efforts of multiple contributors led to the determination of the orbit and stellar parameters for both stars, summarised by \citet{alecian2005}. The latter furthermore present a detailed analysis in terms of stellar calibration of the two individual components in a series of publications \citep{alecian2006, alecian2007b, alecian2007a}. They conclude that RS Cha is composed of two pre-main sequence stars in an evolutionary phase at the first onset of CNO-burning. At this evolutionary stage, central conditions lead to a $\rm{C}^{12}$ depletion by the first steps of the CNO cycle ($\rm{C}^{12}(p,\gamma)\rm{N}^{13}(\beta^+,\nu)\rm{C}^{13}(p,\gamma)\rm{N}^{14}$). The formation of a convective core produces the well known hook in the evolutionary track before the stars' arrival at the zero-age main sequence \citep{iben1965}. Furthermore, the abundance of nitrogen and carbon with respect to the Sun had to be reduced to match the observations \citep{alecian2006, alecian2007b, alecian2007a}.

RS Cha is a circularised and synchronised \citep{alecian2005, woollands2013} binary system of two components of similar mass and luminosity. The orbital and stellar parameters are given in Table \ref{tab:rs_cha_params} and synchronisation is implied by a $v~\mathrm{sin}~i$ ratio equal to the ratio of the radii, the circular orbit \citep{clausen1980} and the expected timescales \citep{zahn1977, mayer1991}. \cite{woollands2013} report a high probability, that is with a $99.9$~\% confidence, that a third component exists within the system, with a mass ranging from $0.3$ to $0.52$ $M_\odot$ and absolute magnitude ranging from $M_v = 24.96$ to $30.32$. This would correspond to the third body being either a white or red dwarf, with the latter being most likely given the evolutionary timescales.

Figure \ref{Fig:1:evolutionary_status} shows the evolutionary status of RS Cha using the result of \citet{alecian2007b}. The coloured lines show stellar evolution tracks that match the observational constraints at a common age of $9.15$~Myr. This is close to the $9.5$~Myr reported by \citet{alecian2007b}. \citet{luhman2004} report an age of $6^\mathrm{+2}_\mathrm{-1}$~Myr that stems from isochrone fitting and is just below this value. The evolution models in Fig. \ref{Fig:1:evolutionary_status} were calculated with version 12778 of the stellar evolution code \textit{Modules for Experiments in Stellar Astrophysics} \citep[\texttt{MESA};][]{paxton2011, paxton2013, paxton2015, paxton2018, paxton2019}.

\begin{figure*}
   \centering
   \includegraphics[width=\linewidth]{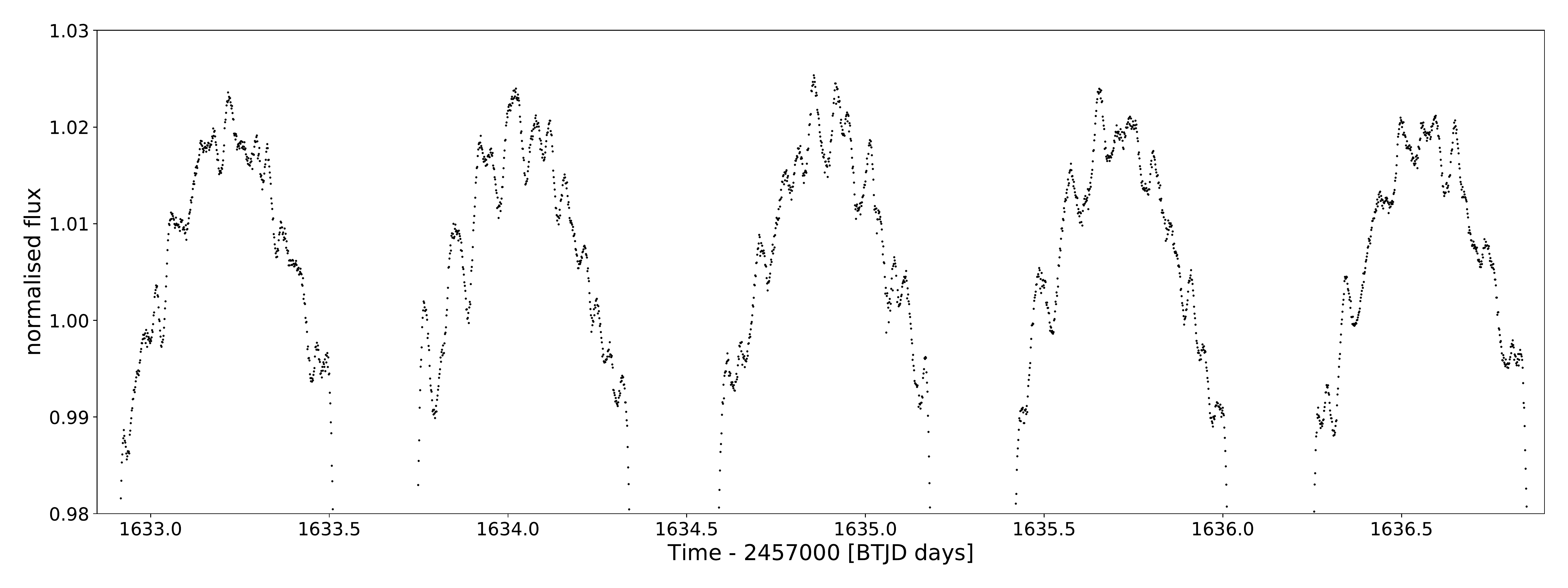}
      \caption{Zoom in on the light curve to make the out-of-eclipse phase variations more visible. The pulsation variations (hours) can clearly be seen.
              }
         \label{Fig:2:zoom_in_lc}
\end{figure*}

\subsection{Pulsations}
Initially observed as part of a programme to detect new $\delta$ Scuti stars, \citet{mcinally1977} were the first to report short-period variability in the light curve of RS Cha. They concluded that at least one component is a $\delta$ Scuti star. \citet{clausen1980} obtained a $22$-d light curve consisting of 1001 individual observations. They found that both components are $\delta$ Scuti stars although they were not able to report a definitive decision. Furthermore, they reported the three most obvious periods: $0.094$~d, $0.086$~d, and $0.072$~d corresponding to frequencies: $10.64$~\cd, $11.63$~\cd, and $13.89$~\cd. 

\citet{alecian2005} investigated radial velocity curves and found regular variations in both components. Because their observations spanned less than $3$~h per night, they were not able to reliably find any pulsation period \citep{alecian2005}. Still, they report $22.079$ \cd as the most prominent variation for the primary component. 

\cite{boehm2009} observed RS Cha for 14 nights with the high-resolution echelle spectrograph Hercules at the Mt John telescope in New Zealand. They report the first clear detection of pulsation frequencies, both for the primary and secondary component. According to their spectroscopic analysis, the primary component shows high degree ($l = 8-11$) pulsation modes with frequencies $21.11$~\cd and $30.38$~\cd while the secondary component shows pulsation modes with $12.81$~\cd ($l = 2~\mathrm{or}~3$), $19.11$~\cd ($l = 0,~10~~\mathrm{or}~13$), and $24.56$~\cd ($l = 6$). 

\section{Observations: TESS data}
\label{sec:observations}
TESS is an MIT-led NASA mission designed for the discovery of planetary transits. The all-sky survey delivers high-quality light curves for more than $200\,000$ stars, rendering it very useful for asteroseismology. RS Cha (TIC 323292655) has been observed by TESS as part of the TASC\_WG04\_SC proposal aiming at studying $\delta$ Scuti stars. The photometric observations of RS Cha embody four sectors, each observed for two spacecraft orbits \citep[i.e. $27.4$~d,][]{ricker2015}. The resulting data product consists of one sector of full frame images (sector 10) and three sectors of short cadence data (sectors 11, 12, 13). For our analysis, we used the short cadence data spanning a total of $84.28$~d. These data are taken with an exposure time of $2$~s and then stacked to produce a cadence of $2$~min.

We accessed the light curve of RS Cha from the Michulski Archive for Space Telescopes (MAST) with the python package \texttt{lightkurve} \citep{barentsen2019}. We used the simple aperture photometry \citep[SAP flux;][]{morris2017}, combined the sectors and normalised the data set according to the median. The SAP flux was preferred over the pre-search data conditioned simple aperture photometry (PDCSAP flux) because the co-trending basis vector analysis introduces instrumental effects especially in sector 12. The resulting light curve is shown in Fig. \ref{Fig:1:Full_LC_p_gaps}, where the eclipsing nature of the binary is clearly visible. The light curve consists of a total of $55\,309$ points with five gaps. These gaps are due to data downlink in between orbits (see red lines in Fig. \ref{Fig:1:Full_LC_p_gaps}) as well as omitted points because of scattered light from the Earth (see blue dashed lines in Fig. \ref{Fig:1:Full_LC_p_gaps}). The gaps have a duration of $2.55$, $1.06$, $1.04$, $1.02$ and $0.93$~d, resulting in a duty cycle of $92.1$~\%. Furthermore, TESS pixels are comparably large and we might expect contamination from nearby sources. The contamination is given as $0.03$, therefore we expect that the data does not suffer from strong third light effects. Figure \ref{Fig:2:zoom_in_lc} shows a zoom in on the first part of sector 12, emphasising the out-of-eclipse light curve. Pulsational variability of the light curve is clearly visible. The strongest modulation is of the order of $2$ to $3$~h, therefore clearly indicating p modes and $\delta$ Scuti{-}type pulsations.

\begin{figure}
   \centering
   \includegraphics[width=\linewidth]{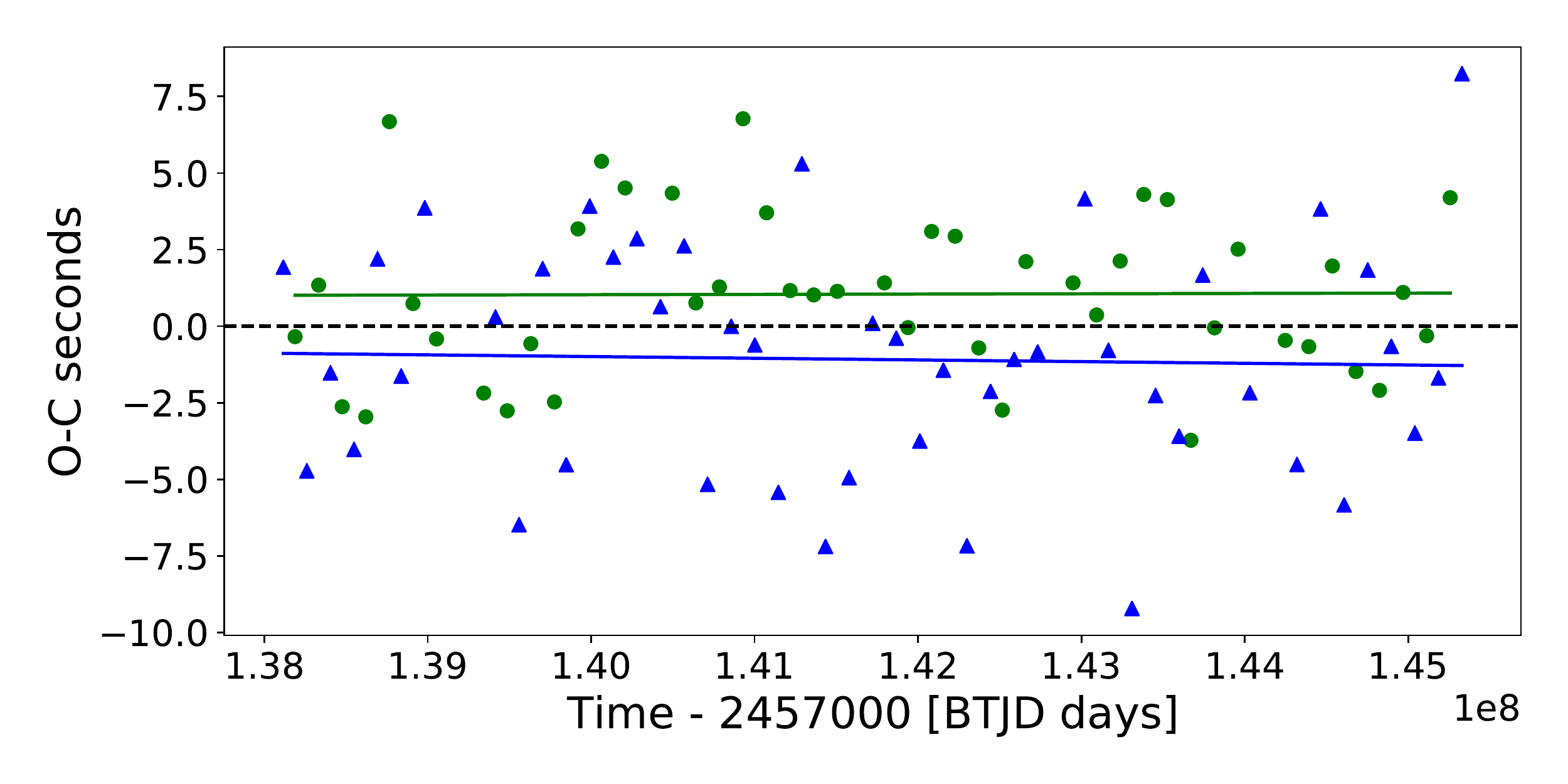}
      \caption{Eclipse timing variations for the light curve after the first iteration. The green circles  and blue triangles show the values for the primary and secondary eclipse respectively. Both data were fitted with a straight line. The small offset to the expected values (black dashed line) could indicate a slight non-zero eccentricity. 
              }
         \label{Fig:3:O_C_diagramm}
\end{figure}

\begin{figure}
   \centering
   \includegraphics[width=\linewidth]{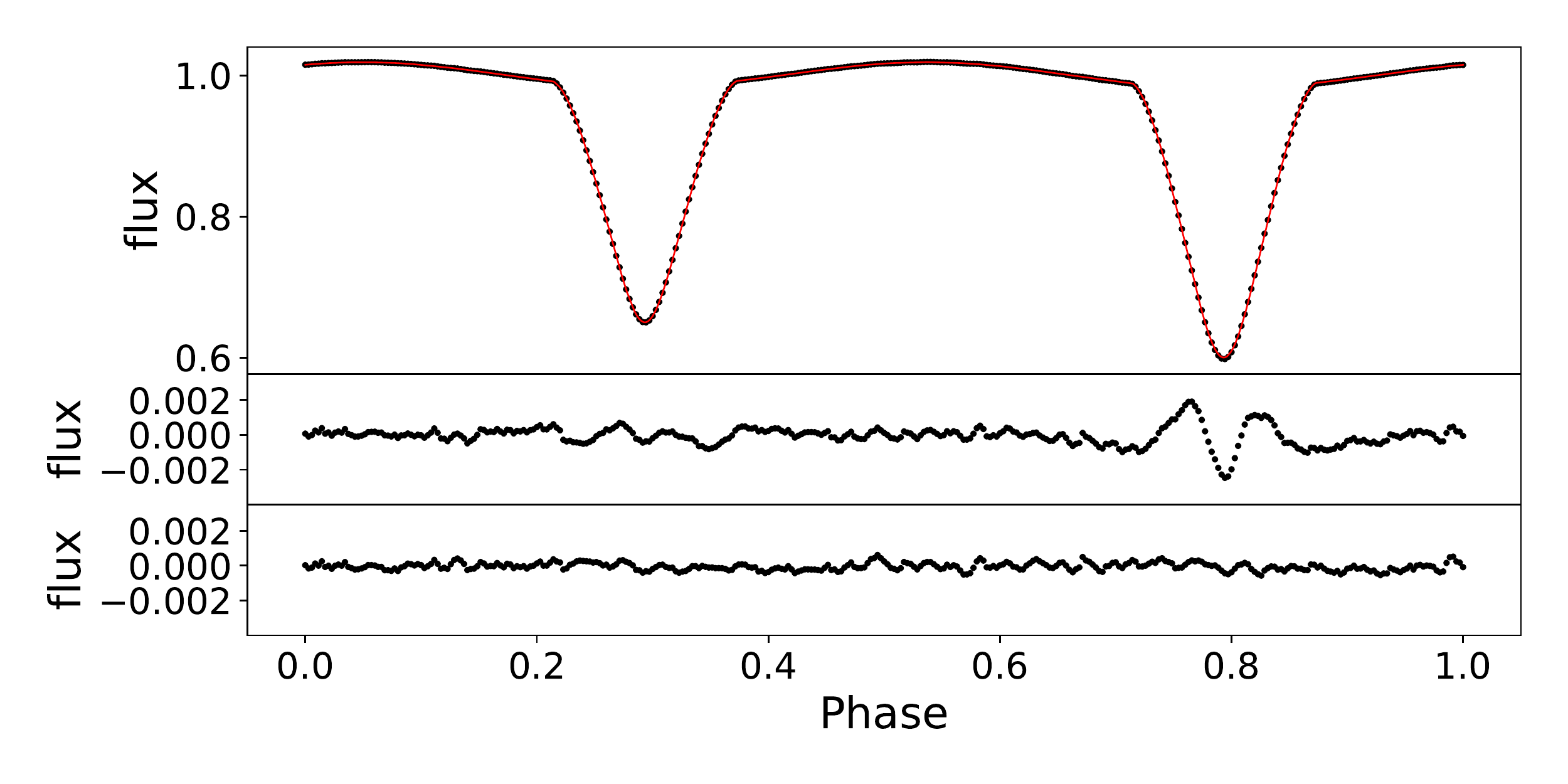}
      \caption{Result of the binary modelling. Top panel: The binned and phase folded observed light curve (black dots) and the final \texttt{PHOEBE} model (red line). Middle panel: The residuals of the best fitting model. Lower panel: The residuals after also removing the fitted orbital harmonics
              }
         \label{Fig:4:Model_residuals}
\end{figure}
  \begin{table}
      \caption[]{Fixed parameters that might be different from \texttt{PHOEBE}'s default binary used in our Nelder-Mead optimisation.
      }
         \label{tab:const_params}
         
    \begin{tabular*}{\linewidth}{l@{\extracolsep{\fill}}r}
            \hline
            \noalign{\smallskip}
            Parameter &  Value  \\
            \noalign{\smallskip}
            \hline
            \noalign{\smallskip}
            orbital period (days)                           & $1.66987725$          \\
            $t_0$~($+ 2457000$~BTJD days)                   & $1599.397275$        \\
            eccentricity ($^\circ$)                         & $0$                   \\
            Synchronicity parameter Primary                 & $1$                   \\
            Synchronicity parameter Secondary               & $1$                   \\
            pblum\_mode                                     & dataset-scaled        \\
            passband                                        & TESS:T                \\
            irrad\_method                                    & horvat \\
            \noalign{\smallskip}
            \hline
    \end{tabular*} 
    \tablefoot{ 
    $t_0$ refers to the time of superior conjunction. 
    Short descriptions of some parameters\footnote{See \url{http://phoebe-project.org/docs/2.2/physics} for more details}: The pblum\_mode defines how the passband intensities are handled, where dataset\_scaled refers to a method that scales the resulting model to best fit the data. The option passband defines which passband is used for the calculations of luminosities, where we use the passband developed for data obtained by TESS. The passband was downloaded from the available online passbands. The option irrad\_method defines which method to use for the handling of irradiation, where we used the Horvat scheme.
    
    }
   \end{table}

\begin{figure*}
   \centering
   \includegraphics[width=\linewidth]{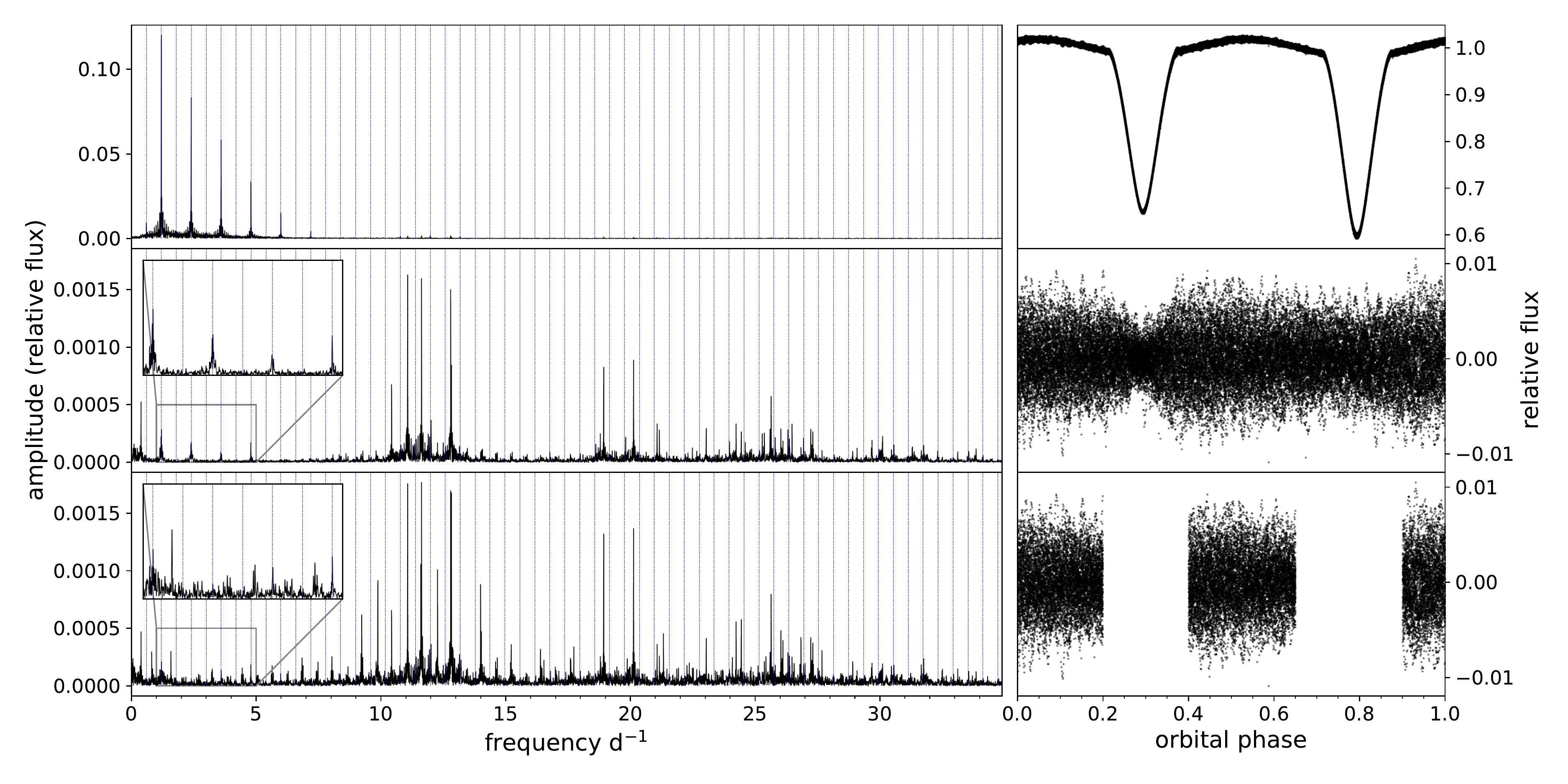}
      \caption{Amplitude spectra and phase folds for different light curves. Top panel: The original light curve. Middle panel: The light curve minus the binary model. Bottom panel: The light curve minus the binary model on the out-of-eclipse phase. The middle and bottom panel have the same scaling in the y-axis while the scaling of the top panel is different.
              }
         \label{Fig:5:Amplitude_spectra}
\end{figure*}

\section{Binary modelling}
\label{sec:binmod}
We use the next generation Wilson-Devinney code \textit{Physics of Eclipsing Binaries} \citep[\texttt{PHOEBE};][]{prsa2005, prsa2016, horvat2018, jones2019} version 2.2.1 to calculate theoretical light curve models for the eclipsing binary. Because the orbital parameters are well constrained (see Table \ref{tab:rs_cha_params}), the aim of the binary modelling is to remove the eclipsing binary signal from the light curve to extract only the pulsational signal.

To do so, we implemented a Nelder-Mead algorithm \citep{nelder1965}, a direct search algorithm often used for optimisation problems with unknown derivatives. The algorithm is further described in Appendix \ref{sec: nelder-mead}. We used an $8-$dimensional parameter space including the mass, equivalent radius, and effective temperature for both the primary and the secondary component as well as the inclination of the system and the time of superior conjunction to optimise a first binary model. The initial values for the mass, equivalent radius, effective temperature, and inclination were taken from Table \ref{tab:rs_cha_params} (references A05 and R00) and we fitted the first primary eclipse with a Gaussian function to obtain an initial value for the time of superior conjunction. The other parameters, which might be different from \texttt{PHOEBE}'s default binary, are kept constant during the iterations and can be found in Table \ref{tab:const_params}. The value for the period was found by applying a phase dispersion minimisation algorithm \citep{stellingwerf1978} on the light curve. We find that aggressive coefficients (see Appendix \ref{sec: nelder-mead}) work best in obtaining good models for the eclipsing light curve after $1000$ iterations. We use a third of the light curve, that is sector 11 and the start of sector 12 to find a good consensus between using most of the available data and execution time. 

The residuals show more than $100$ frequencies that are significant in a classical Fourier analysis, some of which were multiples of the orbital frequency. Because both components are $\delta$ Scuti{-}type stars, a comparable number of frequencies may be expected. The remaining variance at integer multiples of the orbital frequency may be attributed to an inadequate orbital solution because the quality of TESS observations is so good, that the uncertainties are dominated by the choices in the binary modelling. We pre-whitened the light curve for the frequencies that are not multiples of the orbital frequency and applied a Savitzky-Golay filter \citep{savitzky1964} with a window length of $12001$ points to remove irregular low frequency variations before any further optimisation.\footnote{With this window length, the procedure removes intrinsic variability $\lesssim 0.15$~\cd and does not influence the subsequent analysis concerning pulsations.} Such a process is typical when modelling binary stars with pulsating components \citep[see e.g.][]{hambleton2013}. The process consists of multiple iterations until the light curve is thoroughly pre-whitened from any pulsation signal. In this particular case, one iteration of pre-whitening is sufficient.

We used the new binary light curve (the light curve minus the frequencies and the Savitzky-Golay filter) to determine the most precise values for the period and time of superior conjunction by creating an O$-$C diagram. The calculated values for the primary (and secondary) eclipse were obtained via $t_0 + i P$ ($t_0 + (i+\frac{1}{2})P $) where $t_0$ is the time of superior conjunction and $P$ is the period. We optimised both $t_0$ and $P$ to find the most precise parameters. The resulting O$-$C diagram is shown in Fig. \ref{Fig:3:O_C_diagramm}. The best fitting values are a period of $1.66987725(5)$~d and a time of superior conjunction of $1599.397275(5) + 2457000$~BTJD. Thus, these values were fixed in the final optimisation. 

We included gravity darkening and Lambert scattering \citep[(i.e. irradiation method Horvat,][]{prsa2016} in the final Nelder-Mead optimisation. Including the gravity darkening coefficients and the ratio of bolometric light reflected and irradiated by the stars adds four dimensions to the parameter space. All of these values were set to $0.5$. On the other hand, we dropped the time of superior conjunction in the optimisation, leading to a $11$-dimensional parameter space.

Figure \ref{Fig:4:Model_residuals} shows a phase folded and binned version of the full light curve together with the result of the Nelder-Mead optimisation. The residuals show that even with the inclusion of gravity darkening and Lambert scattering the light curve cannot be reproduced perfectly. Therefore, we fitted the remaining residuals with frequencies up to the first $30$ orbital harmonics. The subtraction of this fit yields the residuals in the lower panel of Fig. \ref{Fig:4:Model_residuals}. To obtain the final light curve for the pulsation analysis we calculated a binary model and subtracted it from the full light curve. In addition, the fitted orbital harmonics and a Savitzky-Golay filter were subtracted. The result of this analysis is hereafter called pulsation light curve.
  
\section{Pulsation analysis}
\label{sec:pulsanal}
The light curve resulting from binary modelling demonstrates clear pulsations in the region $10$~\cd\ to $30$~\cd\ as can be seen from the amplitude spectrum (see middle panel of Fig. \ref{Fig:5:Amplitude_spectra}). We used \texttt{smurfs} \citep{mullner2020} and our own codes to generate amplitude spectra and to perform the pulsation analysis. The TESS magnitude of RS Cha is $5.6$. The expected point to point scatter for sectors $11$, $12$, and $13$ is $165$~parts per million (ppm) for short-cadence data at this magnitude. We therefore attribute an average error of this value to the observed flux. In the following, we differentiate between frequencies with high amplitudes ($> 165$~ppm) and frequencies with low amplitudes ($\leq 165$~ppm).

 \begin{table}
    \caption[]{Identified frequencies using superposition of linear modes and their corresponding amplitudes and phases.}
    \label{tab:superp_lin}
    \tabcolsep=0.01cm
    \begin{tabular*}{\linewidth}{lrrrr}
        \hline
        \noalign{\smallskip}
        Designation &\multicolumn{1}{c}{$f$}  &  \multicolumn{1}{c}{$A$}    & \multicolumn{1}{c}{$\phi$} & \multicolumn{1}{c}{$f-if_\mathrm{orb}$}\\
        &\multicolumn{1}{c}{(\cd)} &  \multicolumn{1}{c}{(ppm)}   & \multicolumn{1}{c}{($\frac{\rm rad}{2 \pi}$)}&\multicolumn{1}{c}{(\cd)}\\
        \noalign{\smallskip}
        \hline
        \noalign{\smallskip}
 F1                             & $11.070396(6)$  & $   1671(14)$ & $0.881(6)$ & $0.291161(6)$ \\
 F2                             & $11.624913(6)$  & $   1644(14)$ & $0.951(7)$ & $0.246831(6)$ \\
 F3                             & $12.795260(7)$  & $   1450(14)$ & $0.898(10)$& $0.219486(7)$ \\ 
 F4                             & $20.122555(10)$ & $    909(14)$ & $0.671(12)$& $0.238223(10)$ \\
 F5=F4$-$2$f_\mathrm{orb}$    & $18.924924(11)$ & $    852(14)$ & $0.654(12)$& $0.238161(11)$ \\
 F6=F2+2$f_\mathrm{orb}$    & $12.822585(13)$ & $    751(14)$ & $0.955(16)$  & $0.246810(13)$\\
 F7=F2$-$2$f_\mathrm{orb}$    & $10.427359(13)$ & $    689(14)$ & $0.416(16)$& $0.246970(13)$ \\
 F8                             & $25.636172(16)$ & $    586(14)$ & $0.923(18)$& $0.114224(16)$ \\
 F9                             & $0.384953(17)$  & $    526(14)$ & $0.911(20)$& $0.213894(17)$ \\
 F10                            & $1.20499(9)$    & $    355(6)$  & $0.93(15)$ & $0.00730(9)$ \\
 F11=F2+F9                  & $12.004905(29)$ & $    353(14)$ & $0.14(4)$    & $0.027977(29)$\\
 F12                            & $21.064357(27)$ & $    336(14)$ & $0.96(3)$  & $0.104732(27)$ \\
 F13                            & $24.231017(27)$ & $    333(14)$ & $0.84(3)$  & $0.277161(27)$ \\
 F14=F2$-f_\mathrm{orb}$    & $11.025846(28)$ & $    331(14)$ & $0.17(3)$    & $0.246610(28)$\\
 F15                            & $26.474933(29)$ & $    320(14)$ & $0.46(3)$  & $0.125691(29)$ \\
 F16=2F3                      & $25.58853(3)$   & $    307(14)$ & $0.43(4)$  & $0.16187(3)$ \\
 F17                            & $26.02396(3)$   & $    304(14)$ & $0.26(4)$  & $0.27357(3)$ \\
 F18=F13$-$2$f_\mathrm{orb}$  & $23.03406(3)$   & $    302(14)$ & $0.70(4)$  & $0.27790(3)$ \\
 F19                            & $21.15642(3)$   & $    296(14)$ & $0.86(4)$  & $0.19680(3)$ \\
 F20                            & $18.79018(3)$   & $    283(14)$ & $0.45(4)$  & $0.22594(3)$ \\
 F21                            & $26.30992(3)$   & $    278(14)$ & $0.58(4)$  & $0.03932(3)$ \\
 F22                            & $25.36344(3)$   & $    276(14)$ & $0.93(4)$  & $0.21189(3)$ \\
 F23=F17+2$f_\mathrm{orb}$  & $27.22191(4)$   & $    267(14)$ & $0.76(4)$    & $0.27382(4)$\\
 F24=2F13$-$F19               & $27.30487(4)$   & $    265(14)$ & $0.75(4)$  & $0.24206(4)$ \\
 F25=F15$-$2$f_\mathrm{orb}$  & $25.27759(4)$   & $    261(14)$ & $0.48(4)$  & $0.12604(4)$ \\
 F26=F8$-$2$f_\mathrm{orb}$   & $24.43814(4)$   & $    259(14)$ & $0.56(4)$  & $0.11456(4)$ \\
 F27=2F20$-$F8                & $11.94680(4)$   & $    255(14)$ & $0.99(4)$  & $0.03013(4)$ \\
 F28                            & $19.80501(4)$   & $    235(14)$ & $0.59(4)$  & $0.04308(4)$ \\
 F29=2F10                     & $2.40437(15)$   & $    228(10)$ & $0.53(25)$ & $0.00899(15)$ \\
 F30=F1+3$f_\mathrm{orb}$   & $12.86744(4)$   & $    226(14)$ & $0.68(5)$    & $0.29167(4)$\\
 F31=F10+2F30               & $26.93439(4)$   & $    213(14)$ & $0.69(5)$    & $0.01370(4)$\\
 F32                            & $30.10391(5)$   & $    203(14)$ & $0.89(6)$  & $0.16159(5)$ \\
 F33=F26$-$F27                & $12.49486(5)$   & $    195(14)$ & $0.84(5)$  & $0.08091(5)$ \\
 F34=F33$-f_\mathrm{orb}$   & $11.89432(5)$   & $    193(14)$ & $0.79(6)$    & $0.08261(5)$\\
 F35=F24$-$2$f_\mathrm{orb}$  & $26.10775(5)$   & $    190(14)$ & $0.90(6)$  & $0.24149(5)$ \\
 F36                            & $25.79572(5)$   & $    188(14)$ & $0.45(6)$  & $0.04533(5)$ \\
 F37=F3$-$2F9                 & $12.01945(6)$   & $    184(14)$ & $0.41(7)$  & $0.04252(6)$ \\
 F38                            & $29.67863(6)$   & $    184(14)$ & $0.11(8)$  & $0.26369(6)$ \\
 F39=2F8$-$F24                & $23.96734(5)$   & $    182(14)$ & $0.62(6)$  & $0.01348(5)$ \\
 F40=2F3+2F9                & $26.36555(5)$   & $    182(14)$ & $0.31(6)$    & $0.01631(5)$\\
 F41=F36$-$2$f_\mathrm{orb}$  & $24.59721(5)$   & $    180(14)$ & $0.41(6)$  & $0.04451(5)$ \\
 F42=F26$-$2F2                & $1.19073(9)$    & $    179(18)$ & $0.75(14)$ & $0.00696(9)$ \\
 F43=2F29+F25               & $30.09140(6)$   & $    178(14)$ & $0.69(7)$    & $0.14908(6)$\\
 F44=2F6$-$F26                & $1.21202(16)$   & $    175(6)$  & $0.81(25)$ & $0.01433(16)$ \\
 F45=F8$-$2F7                 & $4.78401(5)$    & $    167(14)$ & $0.45(6)$  & $0.00676(5)$ \\
        \noalign{\smallskip}
        \hline
    \end{tabular*}    
    \tablefoot{
    The values in parentheses give the $1\sigma$ uncertainty as a combination of the values reported by the least square algorithm and the standard error estimates formulated by \citet{montgomery1999}. Here, $f_\mathrm{orb}$ is the orbital frequency. Further possible combinations are: F24=2F15$-$2F6; F33= F22$-$F30; F37=2F12$-$F32; F39=F27+F37; F43=2F26$-$F20; F44=F29$-$F42; F45=F29+2F42 
    }
 \end{table}

\subsection{Pulsation modelling}
\label{sec:linmod}
We model the pulsation light curve using traditional Fourier analysis. The model flux is given by
\begin{equation}
\label{linmod}
    F(t_j) = c + \sum_{i=1}^n A_i \sin\left[2 \pi (f_i t_j + \phi_i)\right]
\end{equation}
for $n$ frequencies, where $A_i$, $f_i$, and $\phi_i$ denote the amplitude, frequency, and phase of the $i$-th frequency, respectively. Furthermore, $c$ is a constant offset that is set to zero for our analysis. The model is evaluated at every observation time $t_j$. Thus, this model has a total of $3n$ free parameters.

The model is produced in an iterative process. In every iteration, a Lomb-Scargle periodogram \citep{lomb1976, scargle1982} is calculated and the frequency and the amplitude of the highest peak is detected. These, and an initial phase of $0.5$, are used to fit a single sine to the residual and the corresponding sine is subtracted from the latter. If the amplitude exceeds four times the local noise level, the frequency is considered to be significant and added to the list of model frequencies \citep{breger1993, kuschnig1997}. At every iteration stage, all frequencies, amplitudes, and phases of the frequencies in the model are fitted to the pulsation light curve.  The iterations were stopped when five consecutive frequencies were insignificant (i.e. S/N $< 4$).

The resulting model has a total of $108$ significant frequencies, $45$ of which show amplitudes higher than $165$~ppm and are specified in Table \ref{tab:superp_lin}. In this set of frequencies, we find two types of combinations: (i) combination with the orbital frequency, and (ii) combinations with other frequencies.  For every frequency $\mathrm{F}_i$ identified by the algorithm we therefore looked for possible combinations of frequencies with higher amplitudes in the form $\mathrm{F}_i = a\mathrm{F}_j + b\mathrm{F}_k$, where $a\in [-2, -1, 1, 2]$ and $b\in [-2, -1, 0, 1, 2]$. A combination was accepted, if the difference was below half of the Rayleigh limit $\frac{1}{T}$, where $T$ is the total length of the light curve \citep{degroote2009}. The identified combinations are shown in Table \ref{tab:superp_lin}. We find $18$ frequencies that are fully independent, mostly in the range of $10-13$ and $18-27$~\cd. The full list of all frequencies extracted, independent of an amplitude cutoff, is shown in Table \ref{tab:appsuperp_lin}.

Table \ref{tab:superp_lin} shows some low frequencies in the range of $0$ to $5$~\cd (F9, F10, F42, F44 and F45). The latter four are close to multiples of the orbital frequency of $0.59884$~\cd. While F42, F44 and F45 are found to be combination frequencies, F10 is independent according to the explanations above. F10 might have been introduced in the binary modelling (see Sect. \ref{sec:binmod}). F9 is unrelated to the orbital frequency and corresponds to a period of $2.598$~days. If the binary system would not be synchronised, this might be a signal of rotational variability of either component. A period of $2.598$~days would correspond to a $v~\mathrm{sin}~i$ of $42(2)$~km $\mathrm{s}^{-1}$ and $46(2)$~km $\mathrm{s}^{-1}$ for the primary and secondary, respectively. This does not fit the observed radial velocity measurements (see Table \ref{tab:rs_cha_params}). We therefore exclude F9 to be a signal of rotational variability.

\begin{figure}
   \centering
   \includegraphics[width=\linewidth]{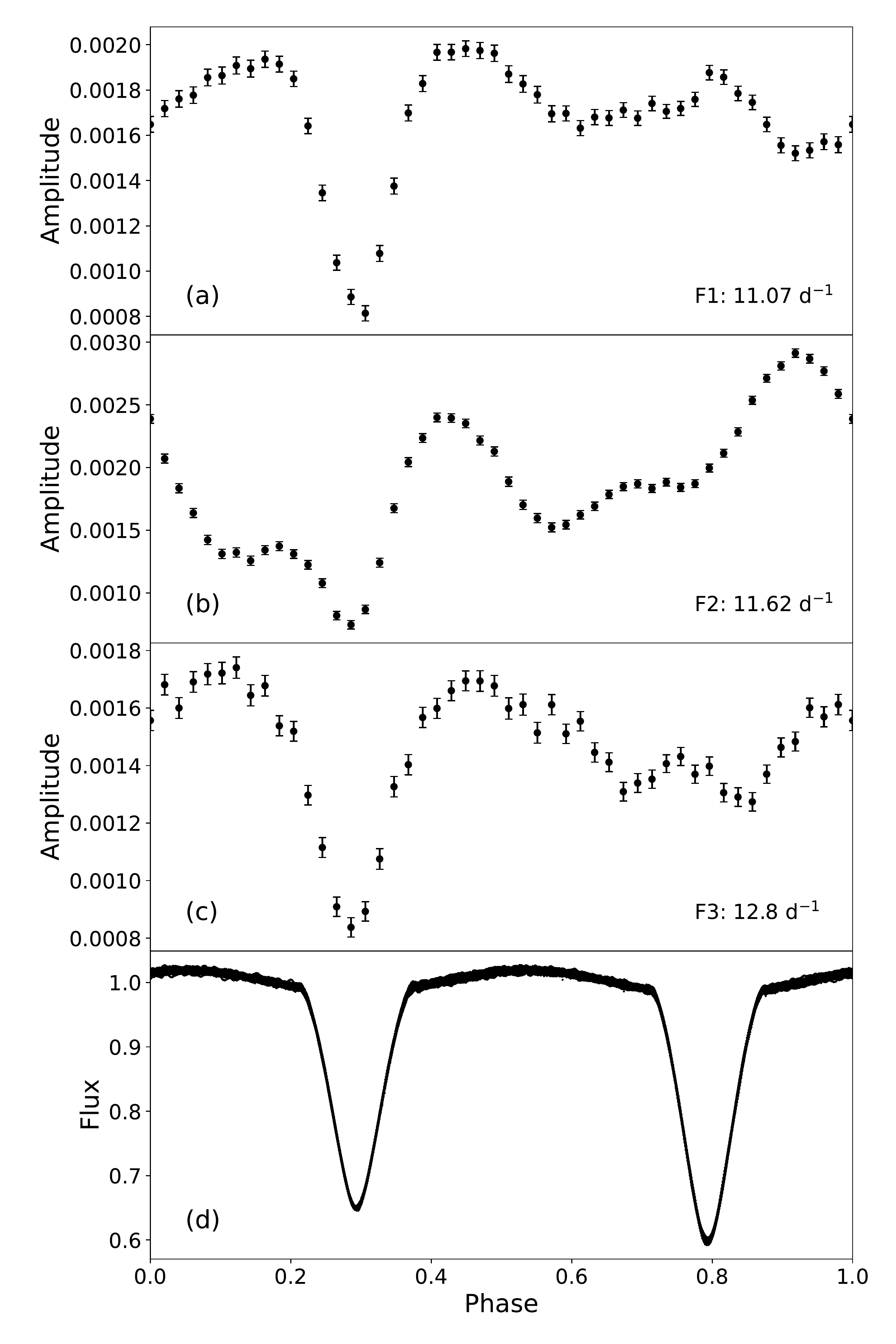}
      \caption{Observed amplitude variations throughout the binary phase. The panels (a), (b), and (c) show the amplitude of the three frequencies with the highest amplitude in the Lomb-Scargle periodogram as a function of the orbital phase, respectively. See the text for additional explanation how the frequencies were fitted. Panel (d) shows the phase folded light curve for comparison. }
         \label{Fig:6:model_motivation}
\end{figure}

\subsection{Binary phase dependent amplitudes}
\label{sec:bpda}
Figure \ref{Fig:5:Amplitude_spectra} compares the amplitude spectra of different versions of the light curve on the left and the corresponding phase folded light curve on the right of each panel. The top panel shows the amplitude spectra of the original light curve, which is dominated by the eclipsing binary signal. The highest peaks are found at two, four, six, eight, and ten times the orbital frequency with amplitudes strongly decreasing towards higher multiples. Furthermore, some small amplitude signal in the range of $10$ to $15$ \cd\ and around $20$ \cd\ is visible. In comparison, the middle panel shows the pulsation light curve which demonstrates clear pulsations especially between $10$~\cd\ and $15$~\cd\ and between $18$~\cd\ and $28$~\cd. The corresponding phase folded light curve shown on the right of the middle panel indicates significant amplitude variations due to the eclipsing nature of RS Cha. During secondary eclipse (and to a smaller extent also during primary eclipse), the pulsation amplitude is clearly diminished. This change in amplitude during eclipses can also be seen in the amplitude spectrum of the out-of-eclipse light curve which is shown in the bottom panel. 

To further investigate this change in amplitude, we fitted the five highest peaks ($11.07$~\cd, $11.62$~\cd, $12.79$~\cd, $20.122$~\cd, $18.92$~\cd) to subsets of the pulsation light curve. The subsets are defined by the orbital phase of the timestamps. We used timestamps with $\phi \pm \Delta \phi$ where $\phi$ is the orbital phase and $\Delta \phi = 0.05$. This subsets include about $5500$ data points each. The uncertainties on the amplitudes is a combination of the uncertainties reported by the least squares fitting algorithm and the standard error estimate $\sigma_A = \sqrt{\frac{2}{N}}\sigma_n$ where $N$ is the number of points in the subset light curve and $\sigma_n$ the root mean square deviation of the observed flux \citep{montgomery1999}. Since this standard error estimate is for data without aliasing problem and the spectral window of this subset light curves is sub-optimal with a duty cycle below $10$\%, we expect these values to be underestimated. The correlation of the measured data points leads to an additional underestimation of the uncertainty \citep{degroote2009}.

The results for the three frequencies with the highest amplitude are shown in Fig. \ref{Fig:6:model_motivation}. The amplitudes of F1 and F3 (top and third panels) show a similar variation: a sharp decrease in amplitude occurs during the secondary eclipse for phases between 0.2 and 0.4, while the amplitudes remain at relatively constant levels during the remaining phases.
This indicates that both frequencies originate from the secondary component. In contrast, F2 shows a very different amplitude modulation: the amplitudes do not remain at relatively constant levels outside of the phases of secondary eclipse, but vary significantly. During secondary eclipse (i.e. between phases from about 0.25 to 0.4), the amplitudes show similar behaviour than for F1 and F3.

\section{Evidence for tidally perturbed modes}
\label{evidence}
The theoretical works by \citet{reyniers2003b, reyniers2003a} and \citet{smeyers2005} discuss the effect of the equilibrium tide on linear, isentropic oscillations of a star in a circular and short-period binary system. The authors predict the presence of perturbed eigenmodes, solidifying themselves as multiplets spaced by (twice) the orbital frequency. A pulsation mode of degree $l$ splits into $l+1$ frequencies corresponding to $|\tilde{m}|= 0, 1, 2, \dots, l$,\footnote{Here, $\tilde{m}$ refers to the azimuthal number in the system of spherical coordinates system where the polar axis aligns with the axis joining the centre of mass of the two stars \citep{reyniers2003a, balona2018}. This is different from the standard azimuthal number $m$ corresponding to the co-rotating coordinate system where the polar axis coincides with the orbital axis.} each of which will further split into multiplets depending on their value of $l$ and $|\tilde{m}|$ respectively. A summary of the theory was given by \citet{balona2018}.

RS Cha is a circularised and synchronised \citep{alecian2005} binary system and is therefore a splendid specimen to test this theory according to which we would see multiple subsets of frequencies spaced by either once or twice the orbital frequency. Furthermore, the recognition of specific frequency patterns may allow the identification of pulsation modes.

\begin{figure}
   \centering
   \includegraphics[width=\linewidth]{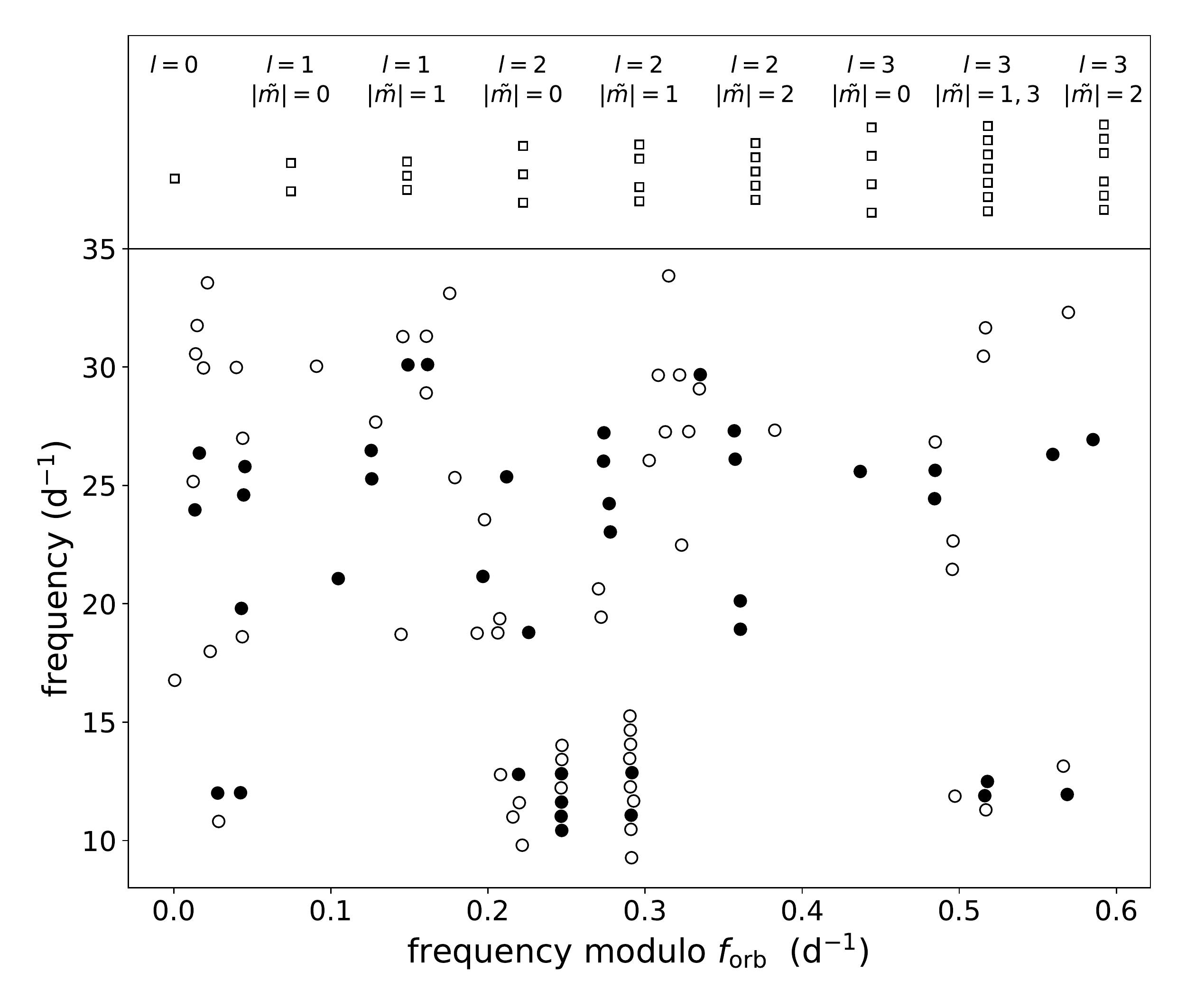}
      \caption{Amplitude spectrum modulo the orbital frequency to visualise the evidence for tidally perturbed modes. Top panel: The signature of multiplets expected from the theory of of perturbed pulsation corresponding to different values of $l$ and $|\tilde{m}|$. Multiplets are plotted according to the tables in \citet{smeyers2005} and \citet{balona2018}. Bottom panel: All significant frequencies from the pulsation analysis above 8 \cd modulo the orbital frequency. Frequencies are shown as black circles where filled circles indicate the frequencies with amplitudes above $165$~ppm and open circles frequencies below that amplitude level. The frequencies are clearly split by the orbital frequencies and often reproduce signatures as expected from the theory.
              }
         \label{Fig:8:perturbed_modes}
\end{figure}

Figure \ref{Fig:8:perturbed_modes} shows all significant frequencies from pulsation modelling nabove 8 \cd modulo the orbital frequency. It is clearly visible that multiple sets of frequencies are split by the orbital frequency. The top panel in Fig. \ref{Fig:8:perturbed_modes} shows the signatures expected from the theory of tidally perturbed pulsations taken from \citet{smeyers2005} and \citet{balona2018}. These signatures can be identified several times in the bottom panel. We therefore see clear evidence for tidally perturbed modes in the light curve of RS Cha. In the following, we will further discuss the observed multiplets and identify the corresponding signatures of tidally perturbed modes.

We start from the frequency with the highest amplitude. F1 is a member of a decuplet (i.e. a multiplet of 10 frequencies, see Table \ref{tab:evidence}). The second highest amplitude component of this decuplet is F30. All other frequencies that are part of this decuplet show low amplitudes. The decuplet cannot be associated with any signature expected from the theory of perturbed oscillations.

F2 is a member of a septuplet of which four frequencies show high amplitudes (see Table \ref{tab:evidence}). Such a septuplet is expected for a $l=3$ $\tilde{m}=1,3$ mode. The second highest amplitude component F6 ($f = 12.822$~\cd) is close to the reported frequency value ($12.81$~\cd) of a $l=2$ or $l=3$ mode \citep{boehm2009}. Dropping two of the low amplitude frequencies (F85 and F86) would lead to a quintuplet, representing the signature of a $l=2$ $\tilde{m}=2$ mode. Together with three low amplitude frequencies, F3 is a member of a quadruplet. Similar to F1, this multiplet cannot be associated with any expected signature. The frequencies F4 and F5 are split by twice the orbital frequency. No other frequencies are associated with this doublet. Thus, its signature therefore corresponds to a $l=1$, $\tilde{m}=0$ mode.

The examples above include only the five highest amplitude frequencies. Beyond them, we can identify multiple further cases which show evidence for tidally perturbed modes. These are given in Table \ref{tab:evidence}. For some of these multiplets dropping a low amplitude frequency leads to alternative mode identification. Since not all frequencies in a multiplet may have a visible amplitude, the amplitude distribution of frequencies within each multiplet is an additional source of uncertainty in the mode identification. 

Tidal perturbation of self excited modes is not the only physical effect that causes frequency multiplets. An additional process that would split the pulsation  frequencies in the amplitude spectrum is amplitude modulation due to arrival time delays \citep{shibahashi2012}. Using the values of Table \ref{tab:rs_cha_params}, we derive the value $\alpha = 0.0092$ resulting in a relative amplitude of the first side peaks of $~\frac{\alpha}{2}=0.0042$ \citep[according to equations (11) and (25) of][]{shibahashi2012}. Higher order side peaks would have even lower amplitudes rendering it impossible to provide that such a rich frequency spectrum as observed in RS Cha originates from this effect.

In Sect. \ref{sec:bpda}, we discussed binary phase dependent amplitudes of the observed frequencies arising from the changing light ratio caused by the eclipses in RS Cha. To investigate the influence of such periodic changes in amplitude, we propose a toy model. In it, we describe the binary phase dependent amplitudes with a Gaussian decrease either during first or secondary eclipse. We discuss the light curves created with these hypothetical amplitude variations and calculate a corresponding pulsation model in Appendix \ref{sec:pdamod}. Indeed, we find frequency multiplets spaced by the orbital frequency in this different model. We then compare the multiplets observed in the toy model (described in Appendix \ref{sec:pdamod}) with the multiplets from the pulsation model described in Sect.
\ref{sec:linmod}. In Table \ref{tab:evidence}, we indicate in the last column if the same multiplet can also be identified in this toy model or not. As such we provide an additional constraint on whether each frequency multiplet is intrinsic to the star, or caused by amplitude modulation during the orbit.

The development of a theoretical description of amplitude variations, similar to the work of \citet{Fuller2020} but including the effects of the eclipsing nature of RS Cha, is needed to further improve the analysis of the pulsation properties of RS Cha. This is however beyond the scope of this work.

 \begin{table}
    \caption{Evidence for tidally perturbed modes and corresponding mode identification.
    }
    \label{tab:evidence}
    \begin{tabular*}{\linewidth}{lccc}
        \hline
        \noalign{\smallskip}
        Frequencies   &  $l$ & $|\tilde{m}|$ & T \\
        \noalign{\smallskip}
        \hline
        \noalign{\smallskip}
         (F1 F30 F51 F72 F79 F82 F88 F92 F102 F105)   & - & - & n\\
         F2 F6 F7 F14 F85 F86 F93                   & 3 & 1,3 & n\\
         F2 F6 F7 F14  F93                          & 2 & 2 & y\\
         (F3 F52 F67 F99)                  & - & - & n\\
         F4 F5                                      & 1 & 0 & y\\
         F8 F26 F54                                 & 2 & 0 & n\\
         F8 F26                                     & 1 & 0 & y\\
         F11 F69                                    & 1 & 0 & y\\
         F13 F18                                    & 1 & 0 & y\\
         F15 F25 F94                                & 2 & 0 & n\\
         F15 F25                                    & 1 & 0 & y\\
         F17 F23                                    & 1 & 0 & y\\
         F24 F35                                    & 1 & 0 & y\\
         F27 F87                                    & 1 & 0 & y\\
         F28 F49                                    & 1 & 0 & y\\
         F32 F57                                    & 1 & 0 & y\\
         F33 F34 F66                                & 1 & 1 & y\\
         F36 F41 F84                                & 2 & 0 & n\\
         F36 F41                                    & 1 & 0 & y\\
         F39 F40 F61                                & 2 & 0 & y\\
         F43 F71                                    & 1 & 0 & y\\
         F47 F50                                    & 1 & 0 & y\\
         F73 F74                                    & 1 & 0 & y\\
         F77 F107                                   & 1 & 0 & y\\
         F98 F101                                   & 1 & 0 & y\\
 
    \noalign{\smallskip}
    \hline
    \end{tabular*}    
    \tablefoot{ Frequencies that are part of multiplets corresponding to evidence of tidally perturbed modes. Following \citet{reyniers2003a, reyniers2003b} and \citet{smeyers2005}, $l$ and $\tilde{m}$ correspond to the mode degree and azimuthal.  The final column T corresponds to whether the same signature can be found in the toy model  of Appendix \ref{sec:pdamod} (y) or not (n). The multiplets around F1 and F3 are in parenthesis, as our analysis suggests that they might originate solely from amplitude variations (see Sect. \ref{pert_vs_var}).}

 \end{table}
 
\subsection{Tidally perturbed modes versus amplitude variations}
\label{pert_vs_var}
Here we further investigate if the observed frequency multiplets in RS Cha could originate solely from amplitude and phase variations over the orbital phase. For this we first pre-whitened the light curve from all pulsation frequencies except those of a corresponding multiplet. To obtain amplitude and phase variations, we fitted the highest amplitude frequency of the multiplet to short subsets of the resulting light curve. Depending on the multiplet, the length of the subsets varies between $0.2$ and $0.3$~days. The values for the amplitude and phase are then folded by the orbital period, binned, and interpolated with a cubic spline to create a template for binary phase dependent amplitudes $A(\theta)$ and phases $\phi(\theta)$, where $\theta$ is the phase of the binary. We then calculate a model flux according to $F_{\rm mod}(t_i) = A(\theta)\sin\left[2 \pi (f t_i + \phi(\theta)\right]$ to see whether such variations can explain the amplitude spectrum obtained from the TESS measurements\footnote{By applying this method to some test cases, we find that any reasonable multiplet in the amplitude spectrum can be obtained to some degree by such variations. This is not surprising as the resulting function phase is not much limited. Hence, by allowing all functional forms for $A(\theta)$ and $\phi(\theta)$, we readily enter the realm of overfitting.} .

Our results of this analysis are provided in Appendix \ref{App:pert_vs_var}. For F1 (see Fig. \ref{F1_amp_variations}), the results are similar to the ones shown in Fig. \ref{Fig:6:model_motivation}: the amplitude is mostly constant but with a decrease during the phase of secondary eclipse. Similarly, the phase is mostly constant out of the eclipse, but shows a phase jump during secondary eclipse. The amplitude spectrum is mostly reproduced by the model flux. These variations can be explained physically: during eclipse, the flux of the secondary is blocked by the primary. With dropping flux levels, the amplitude is decreased. Hence, it is reasonable to assume that the multiplet surrounding F1 is to a large degree created by amplitude variations due to the eclipsing nature of RS Cha. The case of F3 is similar to F1.

On the other hand, for F2 (see Fig. \ref{F2_amp_variations}), the amplitude and phase variations are stronger. The amplitude varies irregularly throughout the orbital phase, but again a decrease during secondary eclipse is observed. Compared to the amplitude variations reported by \citet{Fuller2020}, the amplitude variability of F2 are much more irregular, suggesting that it might be of different origin. The amplitude spectrum is reproduced only qualitatively, with the two second highest peaks showing to small amplitudes. Most importantly, the analysis by \citet{boehm2009} agrees with our mode identification for F2 to be a $l=2$ or $l=3$ mode. Hence, we conclude that the multiplet around F2 is most likely not created by amplitude and phase modulation alone, but rather corresponds to multiple  single frequencies corresponding to tidally perturbed modes as predicted by \citet{reyniers2003a, reyniers2003b}, \citet{smeyers2005} and \citet{balona2018}.

The doublet consisting of F4 and F5 (see Fig. \ref{F4_F5_amp_variations}) represents a third distinct case. The two frequencies are split by twice the orbital frequency, hence the amplitude and phase variations are also dependent on $\theta/2$ rather than $\theta$. In a circularised and synchronised binary system, it remains unclear as to why some modes should show amplitude variability during half the orbital period and others during the full orbital period.
This suggests that the measured amplitude variations are probably just the result of our assumption that the doublet is produced by only one of these frequencies. In addition, the amplitude of the second peak in the amplitude spectrum cannot be reproduced fully.

Other frequency multiplets follow similar behaviour to the doublet consisting of F4 and F5. For lower frequency multiplets, such an analysis is not possible due to the strong scatter in obtained amplitude and phase. This analysis suggests that most likely only the multiplets around F1 and F3 can be the result of single frequencies undergoing amplitude and phase variations, while the others are most likely genuine multiplets. This strengthens the evidence for tidally perturbed pulsations in RS Cha.

\section{Conclusion}
\label{sec:conclusion}
We presented the TESS photometric observations of RS Cha, a pre-main sequence binary system consisting of two $\delta$ Scuti{-}type stars that is circularised and synchronised. Approximately a third of the light curve was modelled with \texttt{PHOEBE} to remove the eclipsing binary signal from the light curve. Because the orbital parameters are well constrained \citep[e.g.][]{alecian2005}, we used a Nelder Mead optimisation to derive a good model for the light curve of the binary system. Some residuals at the time of primary and secondary eclipses were removed by fitting multiples of the orbital frequency to the phase folded residual. 

The remaining residuals were used in the subsequent pulsation analysis. The pulsation modelling resulted in $108$ frequencies, $45$ of which have amplitudes higher than the expected noise level of the corresponding TESS observations ($165$~ppm).

We interpret the results in terms of the theory of tidally perturbed modes according to \citet{reyniers2003b, reyniers2003a} and \citet{smeyers2005}. We find evidence for tidally perturbed modes and use regular patterns in frequency to identify pulsation modes from frequency multiplets. From this analysis, we find that RS Cha seems to mainly pulsate in dipole modes with the addition of one strong $l=2$ or $l=3$ mode. The identification of this multiplet of frequencies agrees to the mode identification based on spectroscopic time series performed by \citet{boehm2009}. 

Three distinct mechanisms lead to frequency multiplets in the amplitude spectrum and therefore influence the extracted frequencies from a classical model of superposition of linear modes. These are: time delay effects \citep{shibahashi2012}, tidally perturbed oscillations \citep{reyniers2003b, reyniers2003a}, and phase dependent amplitudes. While time delay effects can be neglected given the orbital parameters of RS Cha, we investigated the influence of binary phase dependent amplitudes on the observed frequencies with a toy model. From this we conclude that amplitude variations indeed alter the observed frequency multiplets and hence the mode identification. We stress that a full physical model of these amplitude variations is needed to refine the pulsation model of RS Cha in the future. Nonetheless, irrespective of the method used to extract the frequencies, we find frequency multiplets consistent with tidally perturbed pulsation modes.

We investigated the possibility that the multiplets arise from amplitude and phase variations of individual frequencies. We find that only the multiplets around F1 and F3 can be explained by the amplitude variations due to the eclipsing binary nature of RS Cha convincingly. As these two do not correspond to signatures expected from the theory of \citet{reyniers2003a, reyniers2003b} and \citet{smeyers2005}, this strengthens the evidence for tidally perturbed modes in RS Cha.

RS Cha is therefore the first pre-main sequence star that shows tidal effects on its pulsational properties and the first doubly pulsating pre-main sequence star observed with TESS. Such systems are invaluable for the understanding of pre-main sequence asteroseismology. Future efforts will go into the identification and analysis of comparable systems and hence into increasing the sample size. Even though the mode identification can only be used as a first approximation, RS Cha is the first object in which the expected signatures from tidally perturbed modes are used for mode identification. Hence, RS Cha is an ideal specimen to further test the theory of tidally perturbed modes. A full physical model of the amplitude variations, detailed seismic modelling and a quantitative analysis of the theory by \citet{reyniers2003b, reyniers2003a} and \citet{smeyers2005} will be the subject of future work.

Although often overlooked, the pre-main sequence phase plays an important role in stellar evolution --- for example the g-mode frequencies of main sequence stars are strongly influenced by neglecting its pre-main sequence evolution calculations \citep{aerts2018}. Due to the binary nature of RS Cha, the masses and radii of the components are well known. These, together with assumptions of the same age and chemical composition deliver strong constraints for modelling of the stellar interior that allow far more insight than otherwise possible \citep{schmid2016}. This study lays the ground work for subsequent analysis to deliver constraints on angular momentum transport during the pre-main sequence phase. Hence, RS Cha is an important system to enhance our knowledge of stellar structure and evolution in the years to come.

\begin{acknowledgements} 
    We thank Conny Aerts and Bert Pablo for fruitful discussions. D. M. Bowman gratefully acknowledges funding from the European Research Council (ERC) under the European Union’s Horizon 2020 research and innovation programme (grant agreement No. 670519: MAMSIE), and a senior post-doctoral fellowship from the Research Foundation Flanders (FWO) with grant agreement No. 1286521N. \\
    The TESS data presented in this paper were obtained from the Mikulski Archive for Space Telescopes (MAST) at the Space Telescope Science Institute (STScI). Funding for the TESS mission is provided by the NASA Explorer Program. STScI is operated by the Association of Universities for Research in Astronomy, Inc., under NASA contract NAS5-26555. Support for MAST for non-HST data is provided by the NASA Office of Space Science via grant NNX13AC07G and by other grants and contracts. This research has made use of the SIMBAD database, operated at CDS, Strasbourg, France; NASA's Astrophysics Data System; matplotlib, a Python library for publication quality graphics \citep{Hunter:2007}; SciPy \citep{Virtanen_2020}; Astropy, a community-developed core Python package for Astronomy \citep{2018AJ....156..123A, 2013A&A...558A..33A}; NumPy \citep{van2011numpy}; MESA SDK for Mac OS (Version 20.3.1) \citep{townsend2020}.
\end{acknowledgements}

%
%

\bibliographystyle{aa} 
\bibliography{bib} 

\appendix
\section{Nelder-Mead method}
\label{sec: nelder-mead}
The Nelder-Mead method, introduced by \citet{nelder1965} is a simplex method for function minimisation based on a method developed by \citet{spendley1962}. The algorithm we implemented is slightly different from the original Nelder-Mead method and takes the following steps:

At the beginning, an initial simplex of $n+1$ vertices is generated. Here $n$ is the number of free parameters. Each vertex is a point $x_i \in \mathbb{R}^n$ together with a function value $f(x_i)$ where $f: \mathbb{R}^n \rightarrow \mathbb{R}$ is the function to minimise. In our case, $f$ is the $\chi^2$ of the residuals, the binary model flux $F_\mathrm{model}$ subtracted from the light curve flux $F_\mathrm{observed}$,
\begin{align}
    \chi^2 = \sum_{i} (F_{\mathrm{observed}, i} - F_{\mathrm{model}, i})^2 ~.
\end{align}
The simplex is then sorted such that 
\begin{align}
    f(x_1) \leq f(x_2) \leq \cdots \leq f(x_{n+1}) 
\end{align}
and the simplex centroid $x_\mathrm{c}$ for all but the worst vertex is calculated:
\begin{align}
    x_\mathrm{c} = \frac{1}{n}\sum_{k=1}^{n} x_k ~.
\end{align}
At this point the iterations begin.

Since $x_{n+1}$ is the worst point in the simplex, we might expect that the point resulting in a reflection on the centroid is better. Therefore, the reflected point $x_\mathrm{r}$ is calculated:
\begin{align}
    x_\mathrm{r} = (1+\alpha) x_\mathrm{c} - \alpha x_{n+1} ~.
\end{align}
Here $\alpha > 0$ is the reflection coefficient. If the function value of $x_\mathrm{r}$ fulfils $f(x_1) \leq f(x_\mathrm{r}) < f(x_{n})$, the vertex $x_{n+1}$ is replaced by the reflected point and the iteration ends.

If the reflected point is a new minimum, hence $f(x_\mathrm{r}) < f(x_1)$, there might be an even better point further in that direction. Thus, the expanded point $x_\mathrm{e}$ is calculated:
\begin{align}
    x_\mathrm{e} = \gamma x_\mathrm{r}  + (1-\gamma) x_\mathrm{c} ~.
\end{align}
Here, $\gamma > 1$ is the expansion coefficient. The worst point in the simplex is then replaced by either the reflected point or the expanded point, whichever has the lower function value, and the iteration ends.

If on the other hand, the reflected point is worse than $x_n$, we might expect that a vertex better than $x_{n+1}$ lies inside the simplex. Therefore the contracted point $x_\mathrm{t}$ is calculated:
\begin{align}
    x_\mathrm{t} = \beta x_{n+1}  + (1-\beta) x_\mathrm{c} ~.
\end{align}
Here, $\beta$ is the contraction coefficient that lies between $0$ and $1$. If the contracted point is better than the worst point, then the latter is replaced by the contracted point and the iteration ends. Otherwise, the iteration failed. In such a case, the size of the simplex is reduced by replacing all vertices $x_i$ other than the best by
\begin{align}
    x_i = (1-\rho) x_1  + \rho x_i ~.
\end{align}
Here, $\rho$ is the shrink coefficient.

Usual values for the coefficients are $\alpha = 1$, $\gamma = 2$,  $\beta = 0.5$, and $\rho=0.5$. The Nelder-Mead method might be caught in a local minimum. To maximise the chance of escaping such a situation, one can use more aggressive coefficients, $\alpha = 2$, $\gamma = 2$,  $\beta = 0.95$, and $\rho = 0.95$. With such coefficients, the reflected and expanded point lie farther away from the initial simplex, and the size of the simplex is reduced slower, both giving more opportunities to withstand a local minimum.

There are multiple conditions one could apply to terminate the iteration. This includes testing the size of the simplex, the standard deviation of the function values in the simplex, or a pre-specified number of iterations. Experiments show, that the algorithm usually takes $200~n$ iteration to converge towards a final point \footnote{\url{https://www.scilab.org/sites/default/files/neldermead.pdf}},although there is no guarantee that this point is a global minimum! We find that $1000$ iteration suffice, meaning that the best point has not changed for $200+$ iterations. 

As for the termination, there are multiple options on how to build the initial simplex. It should be general enough to allow the algorithm to explore all directions, but also small enough in order not to loose multiple iterations of contraction without improving the best point. We take a given starting point $x_\mathrm{s}$ and create $n$ additional points by moving a small value $\delta$ in either direction of the parameter space
\begin{align}
    x_j = x_\mathrm{s} + \delta_j e_j \quad \quad \mathrm{for} ~~j = 1, \cdots, n ~~.
\end{align}
Here, $\delta_j$ is a small number with $\vert\delta_j\vert \leq 0.05$ and $e_j$ is the unit vector in direction $j$ of our parameter space.

\section{Frequency tables}

\onecolumn

 \begin{table}
    \caption[]{Full list of identified frequencies using superposition of linear modes and their corresponding amplitudes and phases.}
    \label{tab:appsuperp_lin}
    \tabcolsep=0.17cm
    \begin{tabular*}{\linewidth}{lrrr@{\hspace{1.5cm}}lrrr}
        \hline
        \noalign{\smallskip}
        Designation &Frequency  &  Amplitude    & Phase &  Designation &Frequency  &  Amplitude    & Phase \\
        &(\cd) &  (ppm)   & ($\frac{\rm rad}{2 \pi}$) &  &(\cd) &  (ppm)   & ($\frac{\rm rad}{2 \pi}$)\\
        \noalign{\smallskip}
        \hline
        \noalign{\smallskip}
 F1                                 & $11.070396(6)$  & $   1671(14)$ & $0.881(6)$  & 
 F55                  		        & $27.33067(7)$   		& $    131(14)$ 		& $0.14(8)$   \\
 F2                                 & $11.624913(6)$  & $   1644(14)$ & $0.951(7)$  & 
 F56 = F38$-f_\mathrm{orb}$    	& $29.07919(7)$   		& $    131(14)$ 		& $0.42(8)$   \\
 F3                                 & $12.795260(7)$  & $   1450(14)$ & $0.898(10)$ & 
 F57 = F32+2$f_\mathrm{orb}$   	& $31.30082(7)$   		& $    130(14)$ 		& $0.60(8)$   \\
 F4                                 & $20.122555(10)$ & $    909(14)$ & $0.671(12)$ & 
 F58                  		        & $25.33049(7)$   		& $    129(14)$ 		& $0.09(8)$   \\
 F5 = F4$-$2$f_\mathrm{orb}$      & $18.924924(11)$ & $    852(14)$ & $0.654(12)$ & 
 F59 = F42+F42   		            & $2.38469(8)$    		& $    129(14)$ 		& $0.21(10)$  \\
 F6 = F2+2$f_\mathrm{orb}$        & $12.822585(13)$ & $    751(14)$ & $0.955(16)$ & 
 F60 = 2F59+F56   		        & $33.85044(7)$   		& $    126(14)$ 		& $0.42(8)$   \\
 F7 = F2$-$2$f_\mathrm{orb}$      & $10.427359(13)$ & $    689(14)$ & $0.416(16)$ & 
 F61 = F39+2$f_\mathrm{orb}$  	& $25.16393(7)$   		& $    123(14)$ 		& $0.85(9)$   \\
 F8                                 & $25.636172(16)$ & $    586(14)$ & $0.923(18)$ & 
 F62 = 2F44+F23   		        & $29.65189(10)$  		& $    122(15)$ 		& $0.12(14)$  \\
 F9                                 & $0.384953(17)$  & $    526(14)$ & $0.911(20)$ & 
 F63 = F39+4$f_\mathrm{orb}$   	& $23.55283(8)$   		& $    121(14)$ 		& $0.14(9)$   \\
 F10                                & $1.20499(9)$    & $    355(6)$  & $0.93(15)$  & 
 F64 = 2F9+F25    		        & $26.05303(8)$   		& $    120(14)$ 		& $0.82(9)$   \\
 F11 = F2+F9                      & $12.004905(29)$ & $    353(14)$ & $0.14(4)$   & 
 F65 = F24$-$2F48    		        & $22.48058(8)$   		& $    120(14)$ 		& $0.43(9)$   \\
 F12                                & $21.064357(27)$ & $    336(14)$ & $0.96(3)$   & 
 F66 = F33$-$2$f_\mathrm{orb}$    & $11.29616(8)$   		& $    120(14)$ 		& $0.02(9)$   \\
 F13                                & $24.231017(27)$ & $    333(14)$ & $0.84(3)$   & 
 F67 = F3$-$3$f_\mathrm{orb}$   	& $10.99512(8)$   		& $    119(14)$ 		& $0.14(9)$   \\
 F14 = F2$-f_\mathrm{orb}$      & $11.025846(28)$ & $    331(14)$ & $0.17(3)$   &
 F68 = F38$-$F29   		        & $27.27591(9)$   		& $    119(14)$ 		& $0.14(12)$  \\ 
 F15                                & $26.474933(29)$ & $    320(14)$ & $0.46(3)$   & 
 F69 = F11$-$2$f_\mathrm{orb}$    & $10.80785(8)$   		& $    118(14)$ 		& $0.62(9)$   \\
 F16 = 2F3                          & $25.58853(3)$   & $    307(14)$ & $0.43(4)$   & 
 F70 = F47$-$3$f_\mathrm{orb}$   	& $29.96127(8)$   		& $    115(14)$ 		& $0.04(10)$  \\
 F17                                & $26.02396(3)$   & $    304(14)$ & $0.26(4)$   & 
 F71 = F43+2$f_\mathrm{orb}$  	& $31.28586(8)$   		& $    115(14)$ 		& $0.45(10)$  \\
 F18 = F13$-$2$f_\mathrm{orb}$    & $23.03406(3)$   & $    302(14)$ & $0.70(4)$   & 
 F72 = F1+4$f_\mathrm{orb}$    	& $13.46483(8)$   		& $    112(14)$ 		& $0.74(10)$  \\
 F19                                & $21.15642(3)$   & $    296(14)$ & $0.86(4)$   & 
 F73 = F30+F20   		            & $31.65674(8)$   		& $    112(14)$ 		& $0.04(10)$  \\
 F20                                & $18.79018(3)$   & $    283(14)$ & $0.45(4)$   & 
 F74 = F73$-$2$f_\mathrm{orb}$   	& $30.45769(8)$   		& $    111(14)$ 		& $0.32(10)$  \\
 F21                                & $26.30992(3)$   & $    278(14)$ & $0.58(4)$   & 
 F75 = F53+F49   		            & $18.70894(8)$   		& $    110(14)$ 		& $0.78(10)$  \\
 F22                                & $25.36344(3)$   & $    276(14)$ & $0.93(4)$   & 
 F76 = F68$-$F35   		        & $1.16366(9)$    		& $    109(15)$ 		& $0.67(11)$  \\
 F23 = F17+2$f_\mathrm{orb}$      & $27.22191(4)$   & $    267(14)$ & $0.76(4)$   & 
 F77 = F14+F2    		            & $22.65336(8)$   		& $    108(14)$ 		& $0.93(10)$  \\
 F24 = 2F13$-$F19                 & $27.30487(4)$   & $    265(14)$ & $0.75(4)$   & 
 F78 = F19$-$4$f_\mathrm{orb}$   	& $18.75736(10)$  		& $    107(15)$ 		& $0.61(14)$  \\
 F25 = F15$-$2$f_\mathrm{orb}$    & $25.27759(4)$   & $    261(14)$ & $0.48(4)$   & 
 F79 = F1+5$f_\mathrm{orb}$   	& $14.06425(8)$   		& $    107(14)$ 		& $0.49(10)$  \\
 F26 = F8$-$2$f_\mathrm{orb}$     & $24.43814(4)$   & $    259(14)$ & $0.56(4)$   & 
 F80 = 2F23$-$2F1    		        & $32.30832(8)$   		& $    106(14)$ 		& $0.66(10)$  \\
 F27 = 2F20$-$F8                  & $11.94680(4)$   & $    255(14)$ & $0.99(4)$   & 
 F81 = F27+2$f_\mathrm{orb}$  	& $13.14202(9)$   		& $    105(14)$ 		& $0.88(10)$  \\
 F28                                & $19.80501(4)$   & $    235(14)$ & $0.59(4)$   & 
 F82 = F1$-f_\mathrm{orb}$   	& $10.47140(9)$   		& $    104(14)$ 		& $0.98(10)$  \\
 F29 = 2F10                         & $2.40437(15)$   & $    228(10)$ & $0.53(25)$  & 
 F83 = F65+F1    		            & $33.55688(9)$   		& $    102(14)$ 		& $0.41(10)$  \\
 F30 = F1+3$f_\mathrm{orb}$       & $12.86744(4)$   & $    226(14)$ & $0.68(5)$   & 
 F84 = F27+2$f_\mathrm{orb}$  	& $26.99199(9)$   		& $    101(14)$ 		& $0.73(11)$  \\
 F31 = F10+2F30                   & $26.93439(4)$   & $    213(14)$ & $0.69(5)$   & 
 F85 = F2+3$f_\mathrm{orb}$   	& $13.42164(9)$   		& $     99(14)$ 		& $0.27(11)$  \\
 F32                                & $30.10391(5)$   & $    203(14)$ & $0.89(6)$   & 
 F86 = F2+4$f_\mathrm{orb}$   	& $14.02060(9)$   		& $     99(14)$ 		& $0.75(11)$  \\
 F33 = F26$-$F27                  & $12.49486(5)$   & $    195(14)$ & $0.84(5)$   & 
 F87 = 2F48+F61   		        & $29.98219(10)$  		& $     99(14)$ 		& $0.55(12)$  \\
 F34 = F33$-f_\mathrm{orb}$     & $11.89432(5)$   & $    193(14)$ & $0.79(6)$   & 
 F88 = F1$-$3$f_\mathrm{orb}$   	& $9.27411(10)$   		& $     96(14)$ 		& $0.87(11)$  \\
 F35 = F24$-$2$f_\mathrm{orb}$    & $26.10775(5)$   & $    190(14)$ & $0.90(6)$   & 
 F89                 		        & $33.11220(10)$  		& $     96(14)$ 		& $0.25(11)$  \\
 F36                                & $25.79572(5)$   & $    188(14)$ & $0.45(6)$   & 
 F90 = F63$-$F45   		        & $18.77054(12)$  		& $     95(15)$ 		& $0.36(16)$  \\
 F37 = F3$-$2F9                   & $12.01945(6)$   & $    184(14)$ & $0.41(7)$   & 
 F91 = F32$-$2$f_\mathrm{orb}$   	& $28.90534(10)$  		& $     95(14)$ 		& $0.49(11)$  \\
 F38                                & $29.67863(6)$   & $    184(14)$ & $0.11(8)$   & 
 F92 = F1+6$f_\mathrm{orb}$   	& $14.66291(10)$  		& $     94(14)$ 		& $0.59(11)$  \\
 F39 = 2F8$-$F24                  & $23.96734(5)$   & $    182(14)$ & $0.62(6)$   & 
 F93 = F2+$f_\mathrm{orb}$    	& $12.22351(10)$  		& $     94(14)$ 		& $0.61(12)$  \\
 F40 = 2F3+2F9                    & $26.36555(5)$   & $    182(14)$ & $0.31(6)$   & 
 F94 = F15+2$f_\mathrm{orb}$   	& $27.67549(10)$  		& $     94(14)$ 		& $0.65(11)$  \\
 F41 = F36$-$2$f_\mathrm{orb}$    & $24.59721(5)$   & $    180(14)$ & $0.41(6)$   & 
 F95 = F90+$f_\mathrm{orb}$    	& $19.37061(10)$  		& $     93(14)$ 		& $0.17(11)$  \\
 F42 = F26$-$2F2                  & $1.19073(9)$    & $    179(18)$ & $0.75(14)$  & 
 F96 = F68+4$f_\mathrm{orb}$     	& $29.66546(14)$  		& $     93(14)$ 		& $0.01(20)$  \\
 F43 = 2F29+F25                   & $30.09140(6)$   & $    178(14)$ & $0.69(7)$   & 
 F97 = 2F48+F27   		        & $16.76834(10)$  		& $     93(14)$ 		& $0.51(11)$  \\
 F44 = 2F6$-$F26                  & $1.21202(16)$   & $    175(6)$  & $0.81(25)$  & 
 F98 = F74$-$F14   		        & $19.43512(10)$  		& $     92(14)$ 		& $0.54(12)$  \\
 F45 = F8$-$2F7                   & $4.78401(5)$    & $    167(14)$ & $0.45(6)$   & 
 F99 = F3$-$5$f_\mathrm{orb}$   	& $9.80338(10)$   		& $     92(14)$ 		& $0.95(12)$  \\
 F46 = 2F27$-$F37   		        & $11.87532(6)$   & $    161(14)$ & $0.10(7)$   & 
 F100 = F62+F9   		            & $30.03321(10)$  		& $     91(14)$ 		& $0.30(12)$  \\
 F47 = F28+F27   		            & $31.75374(6)$   & $    159(14)$ & $0.77(7)$   & 
 F101 = F98+2$f_\mathrm{orb}$  	& $20.63114(10)$  		& $     90(14)$ 		& $0.70(12)$  \\
 F48 = 2F10   	            	    & $2.40982(22)$   & $   157(10)$  & $0.0(4)$    & 
 F102 = F1+7$f_\mathrm{orb}$  	& $15.26153(10)$  		& $     90(14)$ 		& $0.58(12)$  \\
 F49 = F28$-$2$f_\mathrm{orb}$    & $18.60787(6)$   & $    156(14)$ & $0.26(7)$   & 
 F103 = F62$-$4$f_\mathrm{orb}$  	& $27.26100(12)$  		& $     86(14)$ 		& $0.43(14)$  \\
 F50 = F47$-$2$f_\mathrm{orb}$    & $30.55512(6)$   & $    154(14)$ & $0.41(7)$   & 
 F104 = F59+F42  		            & $3.58133(11)$   		& $     84(14)$ 		& $0.32(12)$  \\
 F51 = F1+2$f_\mathrm{orb}$    	& $12.26757(6)$   & $    153(14)$ & $0.60(7)$   & 
 F105 = F1+$f_\mathrm{orb}$  		& $11.67081(11)$  		& $     83(14)$ 		& $0.73(13)$  \\
 F52 = F3$-$2$f_\mathrm{orb}$     & $11.59805(6)$   & $    150(14)$ & $0.16(7)$   & 
 F106 = F103$-$F88 		        & $17.98861(11)$  		& $     82(14)$ 		& $0.24(13)$  \\
 F53 = 2F30$-$F8    		        & $0.10162(6)$    & $    144(14)$ & $0.94(7)$   & 
 F107 = F77$-$4$f_\mathrm{orb}$   & $21.45520(11)$  		& $     81(14)$ 		& $0.74(13)$  \\
 F54 = F8+2$f_\mathrm{orb}$    	& $26.83395(7)$   & $    136(14)$ & $0.67(8)$   & 
 F108 = F76+F2   		            & $12.78379(13)$  		& $     79(14)$ 		& $0.45(18)$  \\

        \noalign{\smallskip}
        \hline
    \end{tabular*}    
    \tablefoot{ The values in parentheses give the $1\sigma$ uncertainty as a combination of the values reported by the least square algorithm and the standard error estimates formulated by \citet{montgomery1999}. Here, $f_\mathrm{orb}$ is the orbital frequency.}
 \end{table}

 \begin{table}
    \caption[]{Full list of identified frequencies in our toy model with phase dependent amplitude and their corresponding amplitudes and phases.}
    \label{tab:appsuperp_pda}
    
    \tabcolsep=0.12cm
    \begin{tabular*}{\linewidth}{lrrrr@{\hspace{1.4cm}}lrrrr}
        \hline
        \noalign{\smallskip}
        Designation & Frequency  &  Amplitude    & Phase & C & Designation & Frequency  &  Amplitude    & Phase & C \\
        &(\cd) &  (ppm)   & ($\frac{\rm rad}{2 \pi}$) & &  &(\cd) &  (ppm)   & ($\frac{\rm rad}{2 \pi}$) &\\
        \noalign{\smallskip}
        \hline
        \noalign{\smallskip}
 $\tilde{\mathrm{F}}$1                                 & $11.070408(5)$  & $   1766(14)$ & $0.861(6)$  & S &
 $\tilde{\mathrm{F}}$45 = $\tilde{\mathrm{F}}$27$-$2$f_\mathrm{orb}$    & $18.60774(5)$   & $    171(14)$ & $0.48(6)$   & S \\
 $\tilde{\mathrm{F}}$2                                 & $11.624950(6)$  & $   1722(15)$ & $0.891(7)$  & S & 
 $\tilde{\mathrm{F}}$46 = $\tilde{\mathrm{F}}$1+3$f_\mathrm{orb}$       & $12.86735(6)$   & $    167(14)$ & $0.84(7)$   & P \\
 $\tilde{\mathrm{F}}$3                                 & $12.795295(7)$  & $   1533(14)$ & $0.841(9)$  & S & 
 $\tilde{\mathrm{F}}$47 = $\tilde{\mathrm{F}}$27+$\tilde{\mathrm{F}}$24                    & $31.75390(6)$   & $    163(14)$ & $0.51(7)$   & P \\
 $\tilde{\mathrm{F}}$4                                 & $20.122577(10)$ & $    916(14)$ & $0.635(12)$ & S & 
 $\tilde{\mathrm{F}}$48 = 2$\tilde{\mathrm{F}}$46$-\tilde{\mathrm{F}}$8                  & $0.10157(6)$    & $    152(14)$ & $0.02(7)$   & S \\
 $\tilde{\mathrm{F}}$5 = $\tilde{\mathrm{F}}$4$-$2$f_\mathrm{orb}$      & $18.924939(11)$ & $    851(14)$ & $0.629(13)$ & S & 
 $\tilde{\mathrm{F}}$49 = $\tilde{\mathrm{F}}$47$-$2$f_\mathrm{orb}$    & $30.55527(6)$   & $    151(14)$ & $0.17(7)$   & S \\
 $\tilde{\mathrm{F}}$6 = $\tilde{\mathrm{F}}$2$-$2$f_\mathrm{orb}$      & $10.427267(13)$ & $    736(14)$ & $0.560(15)$ & S & 
 $\tilde{\mathrm{F}}$50 = $\tilde{\mathrm{F}}$35+$\tilde{\mathrm{F}}$35                    & $2.38468(6)$    & $    151(14)$ & $0.22(8)$   & S \\
 $\tilde{\mathrm{F}}$7 = $\tilde{\mathrm{F}}$2+2$f_\mathrm{orb}$        & $12.822557(14)$ & $    690(15)$ & $0.006(18)$ & S & 
 $\tilde{\mathrm{F}}$51 = $\tilde{\mathrm{F}}$28+2$f_\mathrm{orb}$      & $31.30091(6)$   & $    150(14)$ & $0.47(7)$   & S \\
 $\tilde{\mathrm{F}}$8                                 & $25.636124(15)$ & $    604(14)$ & $0.001(18)$ & S & 
 $\tilde{\mathrm{F}}$52                                & $27.33145(7)$   & $    135(14)$ & $0.86(8)$   & S \\
 $\tilde{\mathrm{F}}$9                                 & $0.385008(17)$  & $    553(14)$ & $0.821(20)$ & P & 
 $\tilde{\mathrm{F}}$53 = $\tilde{\mathrm{F}}$38+2$f_\mathrm{orb}$      & $31.28600(7)$   & $    134(14)$ & $0.22(8)$   & P \\
 $\tilde{\mathrm{F}}$10 = $\tilde{\mathrm{F}}$9+$\tilde{\mathrm{F}}$2                      & $12.004933(28)$ & $    372(15)$ & $0.09(4)$   & S & 
 $\tilde{\mathrm{F}}$54 = $\tilde{\mathrm{F}}$36+2$f_\mathrm{orb}$      & $25.16413(7)$   & $    134(14)$ & $0.52(8)$   & P \\
 $\tilde{\mathrm{F}}$11                                & $26.474945(26)$ & $    358(15)$ & $0.44(3)$   & P & 
 $\tilde{\mathrm{F}}$55 = $\tilde{\mathrm{F}}$10$-$2$f_\mathrm{orb}$    & $10.80786(7)$   & $    134(14)$ & $0.61(8)$   & S \\
 $\tilde{\mathrm{F}}$12                                & $21.064362(26)$ & $    355(14)$ & $0.95(3)$   & S & 
 $\tilde{\mathrm{F}}$56 = $\tilde{\mathrm{F}}$41 $-\tilde{\mathrm{F}}$35                  & $1.21482(7)$    & $    133(14)$ & $0.21(8)$   & P \\
 $\tilde{\mathrm{F}}$13                                & $24.231101(27)$ & $    343(14)$ & $0.70(3)$   & P & 
 $\tilde{\mathrm{F}}$57 = $\tilde{\mathrm{F}}$28$-$2$\tilde{\mathrm{F}}$50                 & $25.33052(7)$   & $    132(14)$ & $0.05(8)$   & S \\
 $\tilde{\mathrm{F}}$14 = $\tilde{\mathrm{F}}$3+$\tilde{\mathrm{F}}$3                      & $25.588559(29)$ & $    327(14)$ & $0.38(3)$   & P & 
 $\tilde{\mathrm{F}}$58 = 2$\tilde{\mathrm{F}}$56+$\tilde{\mathrm{F}}$25                   & $29.653587)$    & $    131(14)$ & $0.36(9)$   & P \\
 $\tilde{\mathrm{F}}$15                               & $26.024095(29)$ & $    325(14)$ & $0.05(3)$   & S & 
 $\tilde{\mathrm{F}}$59 = $\tilde{\mathrm{F}}$34$-f_\mathrm{orb}$     & $29.07868(7)$   & $    129(14)$ & $0.26(9)$   & P \\
 $\tilde{\mathrm{F}}$16                                & $18.78986(3)$   & $    315(14)$ & $0.98(4)$   & S & 
 $\tilde{\mathrm{F}}$60 = $\tilde{\mathrm{F}}$32$-$2$f_\mathrm{orb}$    & $11.29632(7)$   & $    128(14)$ & $0.75(9)$   & P \\
 $\tilde{\mathrm{F}}$17                                & $1.20448(4)$    & $    311(15)$ & $0.78(5)$   & S & 
 $\tilde{\mathrm{F}}$61 = 2$\tilde{\mathrm{F}}$50+$\tilde{\mathrm{F}}$59                   & $33.85039(7)$   & $    127(14)$ & $0.50(9)$   & S \\
 $\tilde{\mathrm{F}}$18                                & $21.15649(3)$   & $    303(14)$ & $0.76(4)$   & S & 
 $\tilde{\mathrm{F}}$62 = 2$\tilde{\mathrm{F}}$2$-$2$\tilde{\mathrm{F}}$9                  & $22.48058(7)$   & $    127(14)$ & $0.43(9)$   & P \\
 $\tilde{\mathrm{F}}$19 = $\tilde{\mathrm{F}}$11$-$3$f_\mathrm{orb}$    & $25.27767(3)$   & $    301(15)$ & $0.34(4)$   & P & 
 $\tilde{\mathrm{F}}$63 = $\tilde{\mathrm{F}}$18+4$f_\mathrm{orb}$      & $23.55301(7)$   & $    126(14)$ & $0.85(9)$   & S \\
 $\tilde{\mathrm{F}}$20 = $\tilde{\mathrm{F}}$13$-$2$f_\mathrm{orb}$    & $23.03411(3)$   & $    296(14)$ & $0.61(4)$   & S & 
 $\tilde{\mathrm{F}}$64 = 2$\tilde{\mathrm{F}}$9+$\tilde{\mathrm{F}}$19                    & $26.05306(8)$   & $    123(14)$ & $0.77(9)$   & S \\
 $\tilde{\mathrm{F}}$21                                & $26.31002(3)$   & $    285(14)$ & $0.42(4)$   & S & 
 $\tilde{\mathrm{F}}$65 = $\tilde{\mathrm{F}}$26+$\tilde{\mathrm{F}}$2                     & $22.65334(8)$   & $    122(14)$ & $0.97(9)$   & S \\
 $\tilde{\mathrm{F}}$22                                & $25.36352(3)$   & $    277(14)$ & $0.79(4)$   & S & 
 $\tilde{\mathrm{F}}$66 = $\tilde{\mathrm{F}}$48+$\tilde{\mathrm{F}}$45                    & $18.70838(8)$   & $    121(14)$ & $0.71(9)$   & S \\
 $\tilde{\mathrm{F}}$23 = 2$\tilde{\mathrm{F}}$13$-\tilde{\mathrm{F}}$18                 & $27.30485(4)$   & $    271(14)$ & $0.80(4)$   & S & 
 $\tilde{\mathrm{F}}$67 = $\tilde{\mathrm{F}}$46+$\tilde{\mathrm{F}}$16                    & $31.65667(8)$   & $    121(14)$ & $0.17(9)$   & S \\
 $\tilde{\mathrm{F}}$24 = 2$\tilde{\mathrm{F}}$16$-\tilde{\mathrm{F}}$8                  & $11.94672(3)$   & $    270(14)$ & $0.10(4)$   & P & 
 $\tilde{\mathrm{F}}$68 = $\tilde{\mathrm{F}}$34$-\tilde{\mathrm{F}}$41                  & $27.27566(8)$   & $    117(14)$ & $0.55(10)$  & S \\
 $\tilde{\mathrm{F}}$25 = $\tilde{\mathrm{F}}$15+2$f_\mathrm{orb}$      & $27.22180(4)$   & $    257(14)$ & $0.95(4)$   & S & 
 $\tilde{\mathrm{F}}$69 = $\tilde{\mathrm{F}}$18$-$4$f_\mathrm{orb}$    & $18.75596(8)$   & $    115(14)$ & $0.90(10)$  & S \\
 $\tilde{\mathrm{F}}$26 = $\tilde{\mathrm{F}}$2$-f_\mathrm{orb}$      & $11.02619(4)$   & $    257(14)$ & $0.61(4)$   & S & 
 $\tilde{\mathrm{F}}$70 = $\tilde{\mathrm{F}}$62+$\tilde{\mathrm{F}}$1                     & $33.55675(8)$   & $    113(14)$ & $0.61(10)$  & P \\
 $\tilde{\mathrm{F}}$27                                & $19.80505(4)$   & $    254(14)$ & $0.52(4)$   & P & 
 $\tilde{\mathrm{F}}$71 = 2$\tilde{\mathrm{F}}$62$-\tilde{\mathrm{F}}$14                 & $19.37004(9)$   & $    108(14)$ & $0.09(10)$  & P \\
 $\tilde{\mathrm{F}}$28                                & $30.10362(4)$   & $    250(15)$ & $0.36(5)$   & P & 
 $\tilde{\mathrm{F}}$72                                & $33.11220(9)$   & $    106(14)$ & $0.24(10)$  & P \\
 $\tilde{\mathrm{F}}$29 = $\tilde{\mathrm{F}}$22$-$2$f_\mathrm{orb}$    & $24.43831(4)$   & $    241(14)$ & $0.29(4)$   & S & 
 $\tilde{\mathrm{F}}$73 = $\tilde{\mathrm{F}}$17+$\tilde{\mathrm{F}}$17                    & $2.41278(9)$    & $    103(14)$ & $0.15(11)$  & P \\
 $\tilde{\mathrm{F}}$30                                & $26.93403(4)$   & $    216(14)$ & $0.28(5)$   & S & 
 $\tilde{\mathrm{F}}$74 = 2$\tilde{\mathrm{F}}$50+$\tilde{\mathrm{F}}$2                    & $16.39267(9)$   & $    103(14)$ & $0.12(11)$  & S \\
 $\tilde{\mathrm{F}}$31                                & $25.79580(4)$   & $    209(14)$ & $0.32(5)$   & S & 
 $\tilde{\mathrm{F}}$75 = 2$\tilde{\mathrm{F}}$25$-$2$\tilde{\mathrm{F}}$1                 & $32.30819(9)$   & $    102(14)$ & $0.88(11)$  & S \\
 $\tilde{\mathrm{F}}$32 = $\tilde{\mathrm{F}}$29$-\tilde{\mathrm{F}}$24                  & $12.49488(5)$   & $    205(14)$ & $0.81(6)$   & P & 
 $\tilde{\mathrm{F}}$76 = 2$\tilde{\mathrm{F}}$14$-\tilde{\mathrm{F}}$18                 & $30.02491(9)$   & $    100(14)$ & $0.07(11)$  & S \\
 $\tilde{\mathrm{F}}$33 = $\tilde{\mathrm{F}}$31$-$2$f_\mathrm{orb}$    & $24.59732(5)$   & $    202(14)$ & $0.24(5)$   & S & 
 $\tilde{\mathrm{F}}$77 = $\tilde{\mathrm{F}}$68$-\tilde{\mathrm{F}}$39                  & $1.16404(10)$   & $    100(15)$ & $0.03(12)$  & S \\
 $\tilde{\mathrm{F}}$34                                & $29.67910(5)$   & $    198(15)$ & $0.31(6)$   & P & 
 $\tilde{\mathrm{F}}$78 = $\tilde{\mathrm{F}}$3$-f_\mathrm{orb}$      & $12.19408(10)$  & $     99(14)$ & $0.97(12)$  & S \\
 $\tilde{\mathrm{F}}$35 = $\tilde{\mathrm{F}}$29$-$2$\tilde{\mathrm{F}}$2                  & $1.18997(6)$    & $    196(15)$ & $1.00(8)$   & S & 
 $\tilde{\mathrm{F}}$79 = $\tilde{\mathrm{F}}$77+$\tilde{\mathrm{F}}$73                    & $3.58210(9)$    & $     98(14)$ & $0.07(11)$  & S \\
 $\tilde{\mathrm{F}}$36 = 2$\tilde{\mathrm{F}}$8$-\tilde{\mathrm{F}}$23                  & $23.96719(5)$   & $    192(14)$ & $0.86(6)$   & S & 
 $\tilde{\mathrm{F}}$80 = $\tilde{\mathrm{F}}$24+2$f_\mathrm{orb}$      & $13.14177(10)$  & $     98(14)$ & $0.29(11)$  & P \\
 $\tilde{\mathrm{F}}$37 = $\tilde{\mathrm{F}}$36$-\tilde{\mathrm{F}}$24                  & $12.01960(5)$   & $    192(14)$ & $0.16(7)$   & S & 
 $\tilde{\mathrm{F}}$81 = $\tilde{\mathrm{F}}$75$-\tilde{\mathrm{F}}$46                  & $19.43518(10)$  & $     98(14)$ & $0.45(11)$  & P \\
 $\tilde{\mathrm{F}}$38 = 2$\tilde{\mathrm{F}}$29$-\tilde{\mathrm{F}}$16                 & $30.09070(5)$   & $    188(15)$ & $0.85(7)$   & P & 
 $\tilde{\mathrm{F}}$82 = $\tilde{\mathrm{F}}$1+$f_\mathrm{orb}$        & $11.67065(10)$  & $     98(14)$ & $0.12(11)$  & S \\
 $\tilde{\mathrm{F}}$39 = $\tilde{\mathrm{F}}$23$-$2$f_\mathrm{orb}$    & $26.10774(5)$   & $    187(14)$ & $0.91(6)$   & S & 
 $\tilde{\mathrm{F}}$83 = $\tilde{\mathrm{F}}$81+2$f_\mathrm{orb}$      & $20.63132(10)$  & $     95(14)$ & $0.41(12)$  & P \\
 $\tilde{\mathrm{F}}$40 = $\tilde{\mathrm{F}}$32$-f_\mathrm{orb}$     & $11.89424(5)$   & $    186(14)$ & $0.90(6)$   & P & 
 $\tilde{\mathrm{F}}$84 = 2$\tilde{\mathrm{F}}$50+$\tilde{\mathrm{F}}$10                   & $16.76843(10)$  & $     90(14)$ & $0.37(12)$  & P \\
 $\tilde{\mathrm{F}}$41 = 2$\tilde{\mathrm{F}}$17                         & $2.40322(5)$    & $    180(14)$ & $0.46(6)$   & S & 
 $\tilde{\mathrm{F}}$85 = $\tilde{\mathrm{F}}$77+$\tilde{\mathrm{F}}$2                     & $12.78374(12)$  & $     86(14)$ & $0.55(16)$  & S \\
 $\tilde{\mathrm{F}}$42 = $\tilde{\mathrm{F}}$36+4$f_\mathrm{orb}$      & $26.36587(5)$   & $    180(14)$ & $0.78(6)$   & P & 
 $\tilde{\mathrm{F}}$86 = $\tilde{\mathrm{F}}$4$-f_\mathrm{orb}$      & $19.52187(23)$  & $     40(14)$ & $0.85(27)$  & S \\
 $\tilde{\mathrm{F}}$43 = 2$\tilde{\mathrm{F}}$35+$\tilde{\mathrm{F}}$41                   & $4.78393(5)$    & $    177(14)$ & $0.59(6)$   & S & 
 $\tilde{\mathrm{F}}$87 = $\tilde{\mathrm{F}}$2$-$3$f_\mathrm{orb}$     & $9.8310(4)$     & $     26(14)$ & $0.5(4)$    & S \\
 $\tilde{\mathrm{F}}$44 = 2$\tilde{\mathrm{F}}$24$-\tilde{\mathrm{F}}$37                 & $11.87541(5)$   & $    173(14)$ & $0.96(6)$   & S & 
 $\tilde{\mathrm{F}}$88 = $\tilde{\mathrm{F}}$2$-f_\mathrm{orb}$     & $12.2242(4)$    & $     21(15)$ & $0.6(5)$    & S \\
        \noalign{\smallskip}
        \hline
    \end{tabular*}    
    \tablefoot{ The values in parentheses give the $1\sigma$ uncertainty as a combination of the values reported by the least square algorithm and the standard error estimates formulated by \citet{montgomery1999}. Here, $f_\mathrm{orb}$ is the orbital frequency. C is a shortcut for Component and denotes either the primary (P) or secondary (S) whichever was chosen according to the test statistics. From these $88$ frequencies, $30$ were assigned to the primary and $58$ to the secondary component.}
 \end{table}
 
\twocolumn

\section{Superposition of modes with phase dependent amplitudes}
 
\label{sec:pdamod}
\begin{figure}
   \centering
   \includegraphics[width=\linewidth]{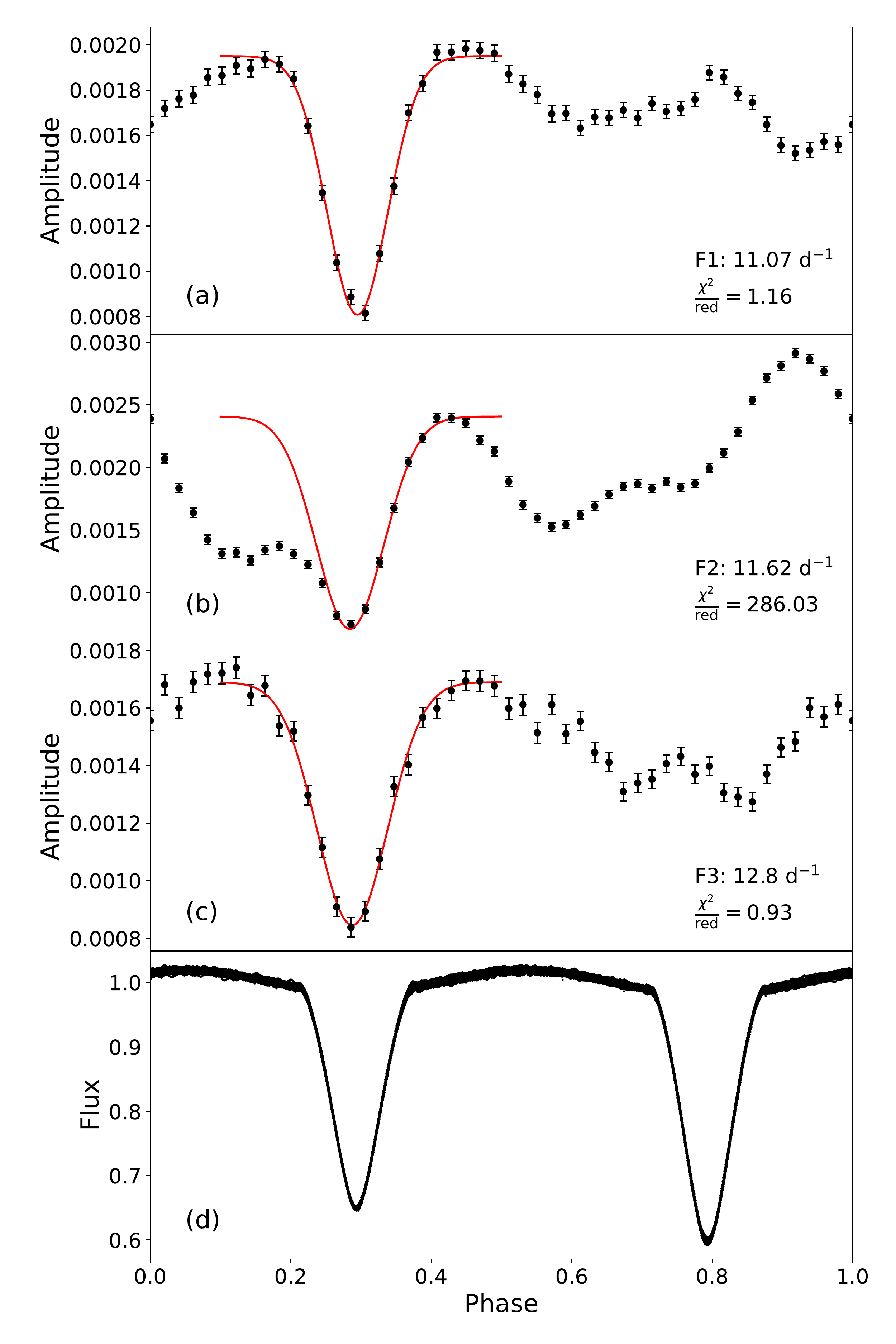}
      \caption{Same as Fig. \ref{Fig:6:model_motivation} but including fits of equation \eqref{pdamodsingle} as red lines. The values for the frequencies F1 and F3 have been fitted between the phases $0.1$ and $0.5$ while the values for F2 have been fitted between $0.25$ and $0.45$. The reduced  chi-square  statistic  is  given  in  the  bottom  right  of  each  panel indicating a good fit for F1 and F3.
              }
         \label{model_mot}
\end{figure}

To investigate the influence of binary phase dependent amplitudes we propose a toy model, in which we assume a decrease in amplitude described by a negative Gaussian peak at the time of eclipse. Figure \ref{model_mot} shows the amplitude variations of the three highest amplitude frequencies (same values as in Fig. \ref{Fig:6:model_motivation}). We fitted such a negative Gaussian peak to the values around secondary eclipse and find good agreement for F1 and F3. Albeit F2 shows good agreement in its respective fitting area ($0.25 \leq \phi \leq 0.45$) the poor agreement elsewhere leads to a reduced chi-squared statistic of $\frac{\chi^2}{\rm red} = 286.03$. The origin of the amplitude variation of F2 outside secondary eclipse is unclear. Figure \ref{model_mot} shows that that the amplitude variations during the eclipse can mostly be described by a Gaussian model and hence suggests that we will obtain insight on the influence of amplitude variations on the extracted multiplets.

In this model, any frequency is assigned to either the primary or secondary star. The model flux is given by
\begin{equation}
\label{pdamod}
\begin{split}
    F_{\rm toy}(t_j) =& c + \sum_{i=1}^{n_\mathrm{p}} A_i \,D(t_j, \delta_\mathrm{p}, \phi_\mathrm{0,p}, \sigma_\mathrm{p}) \sin\left[2 \pi (f_i t_j + \phi_i)\right] +\\
    &+\sum_{k=1}^{n_\mathrm{s}} A_k \,D(t_j, \delta_\mathrm{s}, \phi_\mathrm{0,s}, \sigma_\mathrm{s}) \sin\left[2 \pi (f_k t_j + \phi_k)\right],
\end{split}
\end{equation}
where
\begin{equation}
\label{pdamodsingle}
    D(t, \delta, \phi_\mathrm{0}, \sigma)  = 1-\delta\exp\left[-\left(\frac{t \,\mathrm{modulo}\, P}{P}-\phi_\mathrm{0}\right)^2/(2\sigma^2)\right] \, ,
\end{equation}
describes the decrease in amplitude during eclipses. Here, $n_\mathrm{p}$ ($n_\mathrm{s}$) is the number of frequencies assigned to the primary (secondary) component, the meaning of $c$ is as in equation \eqref{linmod}, $\delta$ describes the amount of the amplitude decrease (i.e. the defect), $\phi_\mathrm{0}$ the orbital phase of the minimum, $\sigma$ the standard deviation of the Gaussian and $P=1.66987725$~d the orbital period of the system. As in equation \eqref{linmod}, $A_i$, $f_i$, and $\phi_i$ denote the amplitude, frequency and phase of the $i$-th frequency respectively. The subscripts p and s correspond to values for the primary and secondary. 

The model has a total number of $3n_\mathrm{p}+3n_\mathrm{s} + (3)_\mathrm{p}+(3)_\mathrm{s}$ free parameters, where the parentheses denote that this numbers are only added if there is at least one frequency assigned to the primary or secondary component respectively.

\begin{figure}
   \centering
   \includegraphics[width=\linewidth]{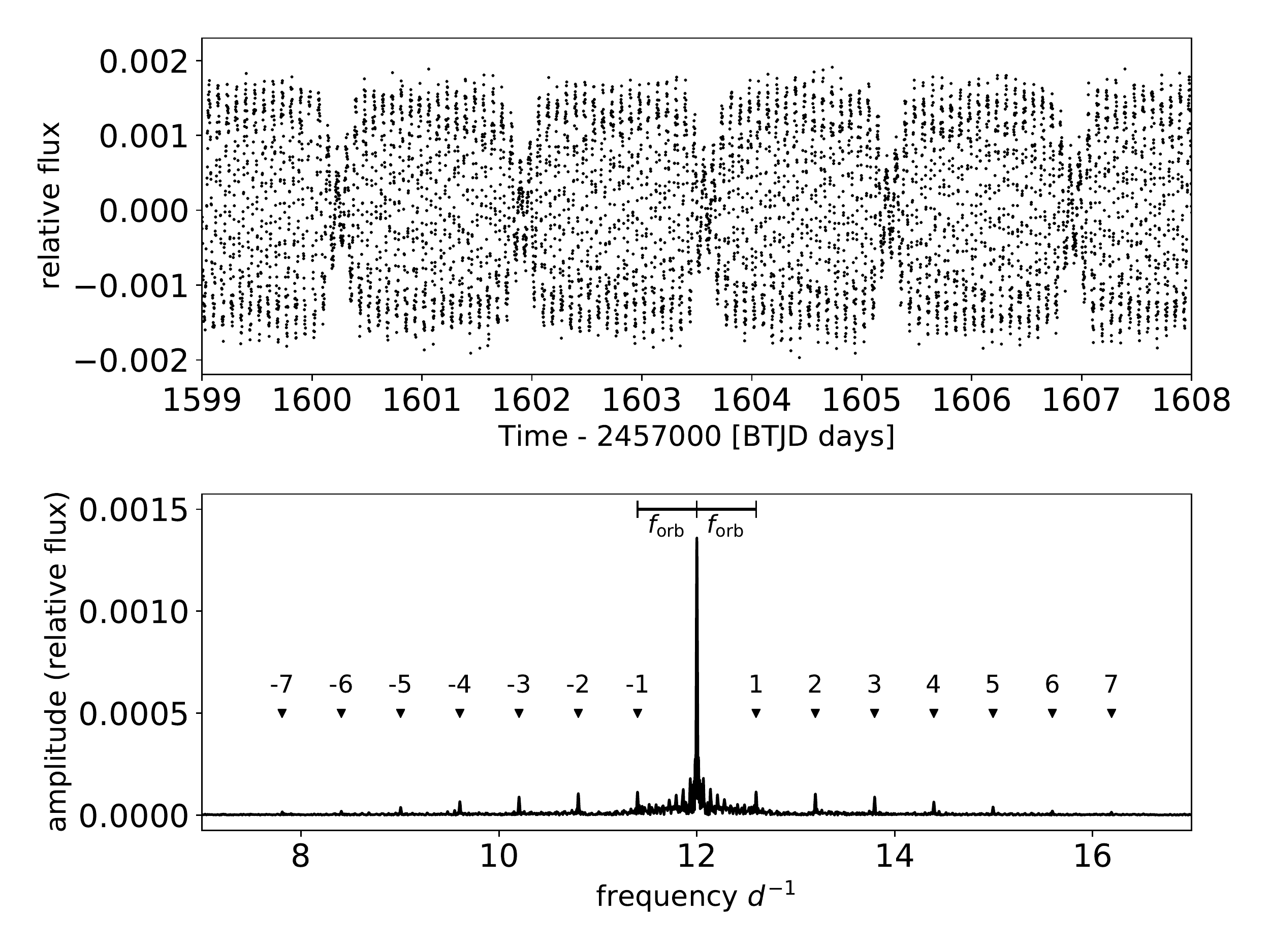}
      \caption{Top panel: simulated light curve for one frequency originating from the secondary component. Bottom panel: amplitude spectra of the light curve in the top panel. Side-frequencies are marked with a triangle and annotated with the number of orbital frequencies that they are shifted with. The errorbars on top show the extend of one orbital frequency. 
              }
         \label{App:1:lc}
\end{figure}
To show the effect of the phase dependent amplitude as in equation \eqref{pdamod}, we simulated a light curve for a frequency of $12$ \cd\ with an amplitude of $1500$~ppm for the secondary star for an amplitude decrease of $0.7$ and a standard deviation of the Gaussian of $0.05$. The results are shown in Fig. \ref{App:1:lc}. The amplitude of the signal is clearly modulated during the time of secondary eclipse. This results in a splitting of the signal in the amplitude spectra, where the peaks are distanced by $i f_\mathrm{orb}$ with $i \in [-7, -6, -5, -4, -3, -2, -1, 1, 2, 3, 4, 5, 6, 7]$. All of this peaks are also extracted from a modelling process according to Equation \eqref{linmod}.

\begin{figure}
   \centering
   \includegraphics[width=\linewidth]{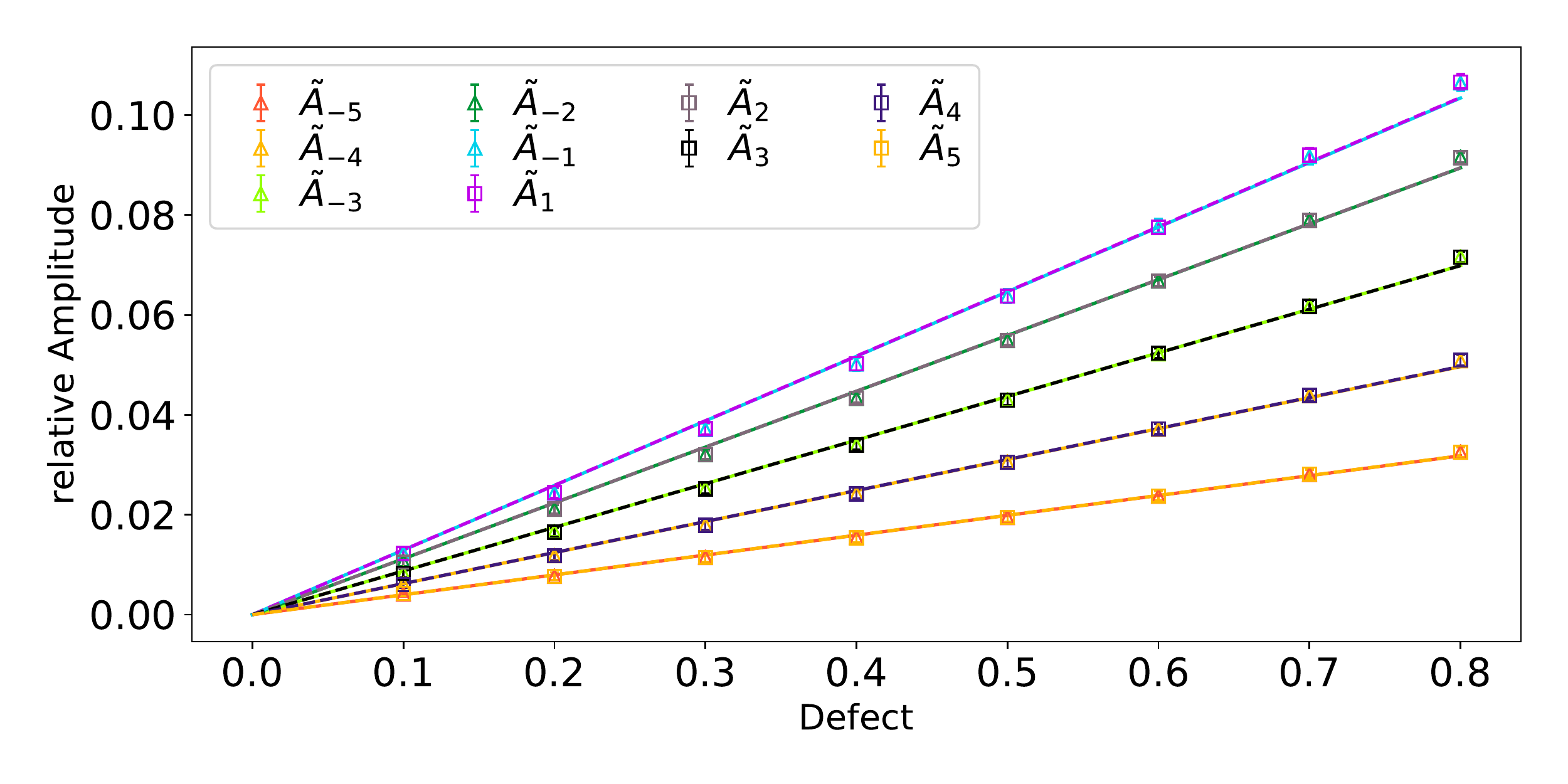}
      \caption{Relative amplitude of the side peaks as a function of the defect where different colours correspond to different side peaks as given by the legend. The uncertainties are typically the size of the markers. Lines correspond to linear fits that agree within $2 \sigma$ with a reduced chi-squared statistics of $1$ for all fits.
              }
         \label{App:2:stats}
\end{figure}

We furthermore simulated $1000$ light curves for different amplitude decrease between $0.1$ and $0.8$ in steps of $0.1$ with different frequencies in the range $10 \leq f \leq 20$ \cd\ and amplitudes in the range $1000 \leq \Tilde{A} \leq 3000$~ppm. We modelled all these light curves using equation \eqref{linmod} and analyse the results. For all the light curves, we found side peaks corresponding to $|i| \leq 5$. The corresponding amplitudes relative to the input amplitudes are shown in Fig. \ref{App:2:stats} and denoted with $\Tilde{A}_i$. The relative amplitude of the sidepeaks is in a first approximation linearly dependent on the defect (i.e. the depth of the Gaussian decrease) and symmetric, that is $\Tilde{A}_i = \Tilde{A}_{-i}$. Fitting the data points with a polynomial of degree two, the values become overfitted according to the reduced chi-squared statistics.

\begin{table}
    \caption[]{Parameters for the phase dependent amplitudes.}
    \label{tab:pda_overall}
    
    \begin{tabular*}{\linewidth}{lrrrr}

        \hline
        \noalign{\smallskip}
        Component & \multicolumn{1}{c}{$\phi_0$}  & \multicolumn{1}{c}{$\sigma$}     &\multicolumn{1}{c}{$\delta$}  &\multicolumn{1}{c}{\#}  \\
        &\multicolumn{1}{c}{($\frac{\rm rad}{2 \pi}$)} & \multicolumn{1}{c}{($\frac{\rm rad}{2 \pi}$) }   &  &\\
        \noalign{\smallskip}
        \hline
        \noalign{\smallskip} 
        Primary     & $0.7913(4)$   & $0.0481(4) $  & $0.511(3)$    & 14 \\
        Secondary   & $0.29098(8)$  & $0.02768(9) $ & $0.6470(16)$  & 32\\
        \hline
        \noalign{\smallskip}
    \end{tabular*}    
    \tablefoot{
    The values in parentheses give the $1\sigma$ uncertainty as reported by the least square algorithm. \# denotes the number of frequencies with amplitudes higher than 165~ppm identified for each component.}

 \end{table}
 
Similar to the superposition of linear modes which we call our traditional method, this new model is produced in an iterative process. An initial guess is extracted from the amplitude spectrum of the residuals and fitted to the residual light curve with the function for phase dependent amplitude
\begin{equation}
    \mathrm{PDA}(t_j) = A \,D(t_j, \delta, \phi_\mathrm{0, p/s}, \sigma) \sin\left[2 \pi (f t_j + \phi)\right] \, .
\end{equation}
The initial values and constraints for $\delta$ and $ \sigma$ are $0.75\pm0.15$ and $0.05\pm0.025$, respectively, and the fit is produced for a primary and a secondary model taking the values and constraints $\phi_\mathrm{0, p} = 0.8\pm 0.05$ and  $\phi_\mathrm{0, s} = 0.3\pm 0.05$, respectively. We then calculate the variance reduction (VR), power reduction (PR), Akaike's information criterion (AIC) and the Bayesian information criterion (BIC) for both of the fits and use these statistics to assign the frequency to a star. A description of these criteria can be found in Sect. 3.4 of \citet{degroote2009}. In the assignment, the AIC and the BIC are the preferred statistics. The frequency is again only added to the model if the amplitude exceeds four times the local noise level \citep{breger1993, kuschnig1997}.  The iterations were stopped when five consecutive frequencies were insignificant (i.e. S/N $< 4$). At every iteration stage all frequencies, amplitudes, and phases for the frequencies in the model as well as the parameters for the phase dependent amplitude for the primary and secondary are fitted to the pulsation light curve with an iterative approach. First the modes of both stars are fitted to the pulsation light curve separately. Then, for each iteration, a model of the pulsating secondary component is subtracted from the pulsation light curve, the modes of the primary star fitted to the residuals and vice versa. This separate fitting and subtraction from the pulsation light curve is repeated an arbitrary five times.

The parameters for the phase dependent amplitudes in this approach are given in Table \ref{tab:pda_overall}. In a circularised orbit, the expected phase difference between primary and secondary eclipse is equal to $0.5$~$\frac{\rm rad}{2 \pi}$. The difference between the phase of the minimum for the primary and secondary follows this expectation within $1\sigma$. The resulting model has a total of $88$ significant frequencies, $46$ of which show amplitudes higher than $165$~ppm. The latter split in $32$ assigned to the secondary and $14$ assigned to the primary component. Therefore, this model has $20$ low amplitude frequencies less than the traditional model, which has $108$ significant frequencies and $45$ with amplitudes higher than $165$~ppm. Table \ref{tab:appsuperp_pda} presents all frequencies in our different approach. We find similar combinations as for the traditional model (see Table \ref{tab:appsuperp_lin}). The designation to a component is certainly not foolproof, especially for lower amplitude modes, but should be plausible for higher amplitude modes ($A \gtrsim 500$~ppm). Indeed, it is a reassuring sign that $\tilde{\mathrm{F}}$5, $\tilde{\mathrm{F}}$6 and $\tilde{\mathrm{F}}$7 all point to the secondary component as their parent frequencies $\tilde{\mathrm{F}}$2 and $\tilde{\mathrm{F}}$4. This relationship does not hold for all frequencies with amplitudes $\lesssim 500$~ppm, however. Similar to the traditional model, the toy model shows $18$ fully independent frequencies that coincide. We find that the frequencies in Table \ref{tab:appsuperp_pda} and Table \ref{tab:appsuperp_lin} resemble each other very well, apart from the amplitude.

\subsection{Model comparison}
\label{sec:mcomp}

\begin{figure}
   \centering
   \includegraphics[width=\linewidth]{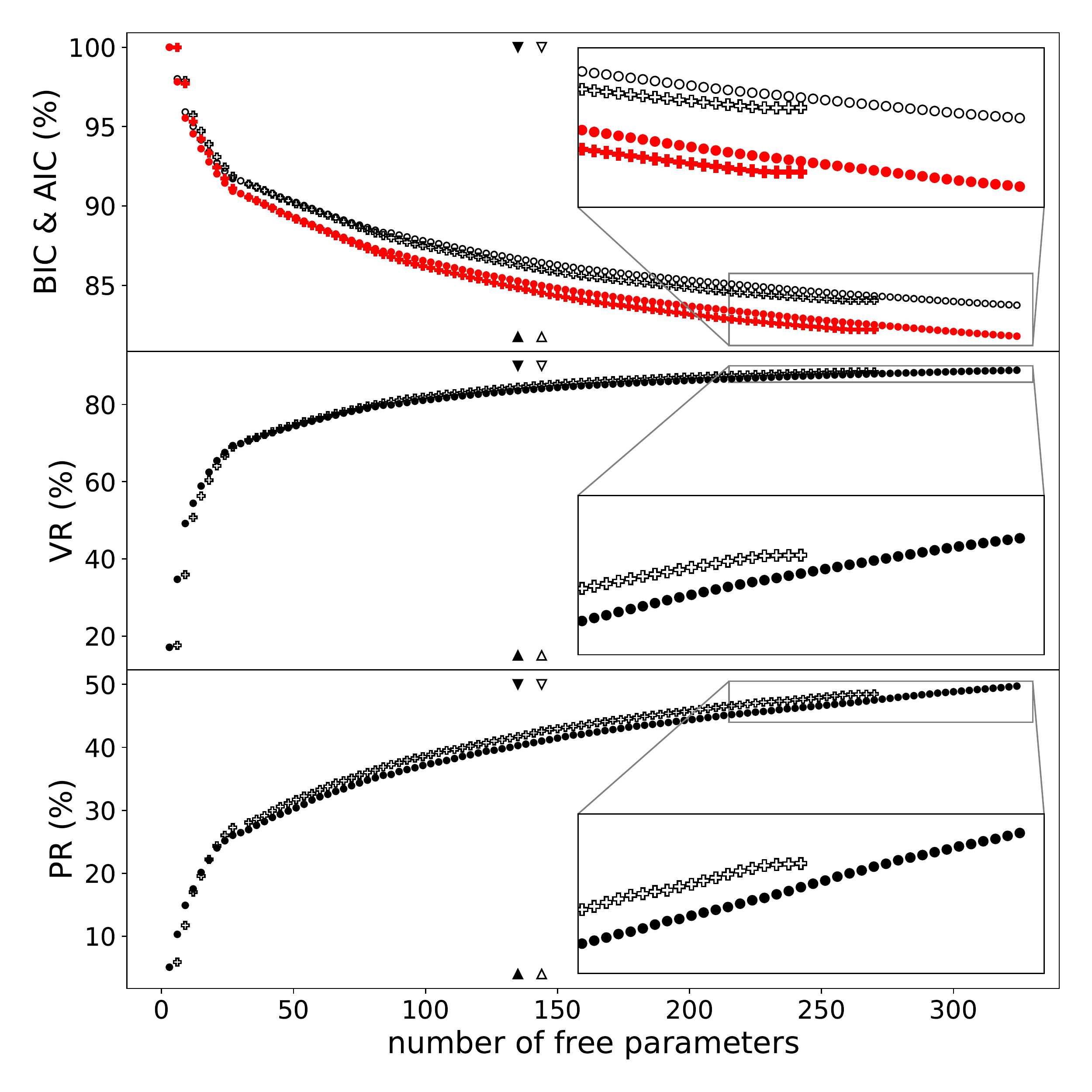}
      \caption{Four different test statistics as a function of the number of free parameters. Circles denote the traditional model while pluses denote the toy model .The filled (open) triangles mark the number of free parameters using only the frequencies with amplitude higher than $165$~ppm for the traditional model (toy model). Top panel: AIC and BIC, where black open symbols correspond to the BIC and red filled symbols to the AIC. These statistics were calculated in respect to pulsation light curve. Middle panel: VR. Bottom panel: PR. All the insets show zoom-ins towards the areas concerning higher number of parameters.
              }
         \label{Fig:7:model_statistics}
\end{figure}

Figure \ref{Fig:7:model_statistics} shows the calculated model statistics, where the top panel compares the AIC and BIC for, the middle panel shows the VR and the lower panel shows the PR of both models. All statistics agree on the following: the model with phase dependent amplitudes describes the light curve better at a fixed number of degrees of freedoms (and therefore model frequencies). The additional frequencies in the traditional model although improve this model further, outweighing the cost of additional free parameters for both the AIC and BIC. Using only the frequencies with amplitude higher than $165$~ppm, all statistics agree that modes with phase dependent amplitude model the pulsation light curve of RS Cha more successfully than classical linear modes. Furthermore, the BIC and AIC discourages the use of the last three frequencies for the toy model, which is not surprising, given their low amplitudes. 

 \subsection{Influence on frequency multiplets}
\begin{figure}
   \centering
   \includegraphics[width=\linewidth]{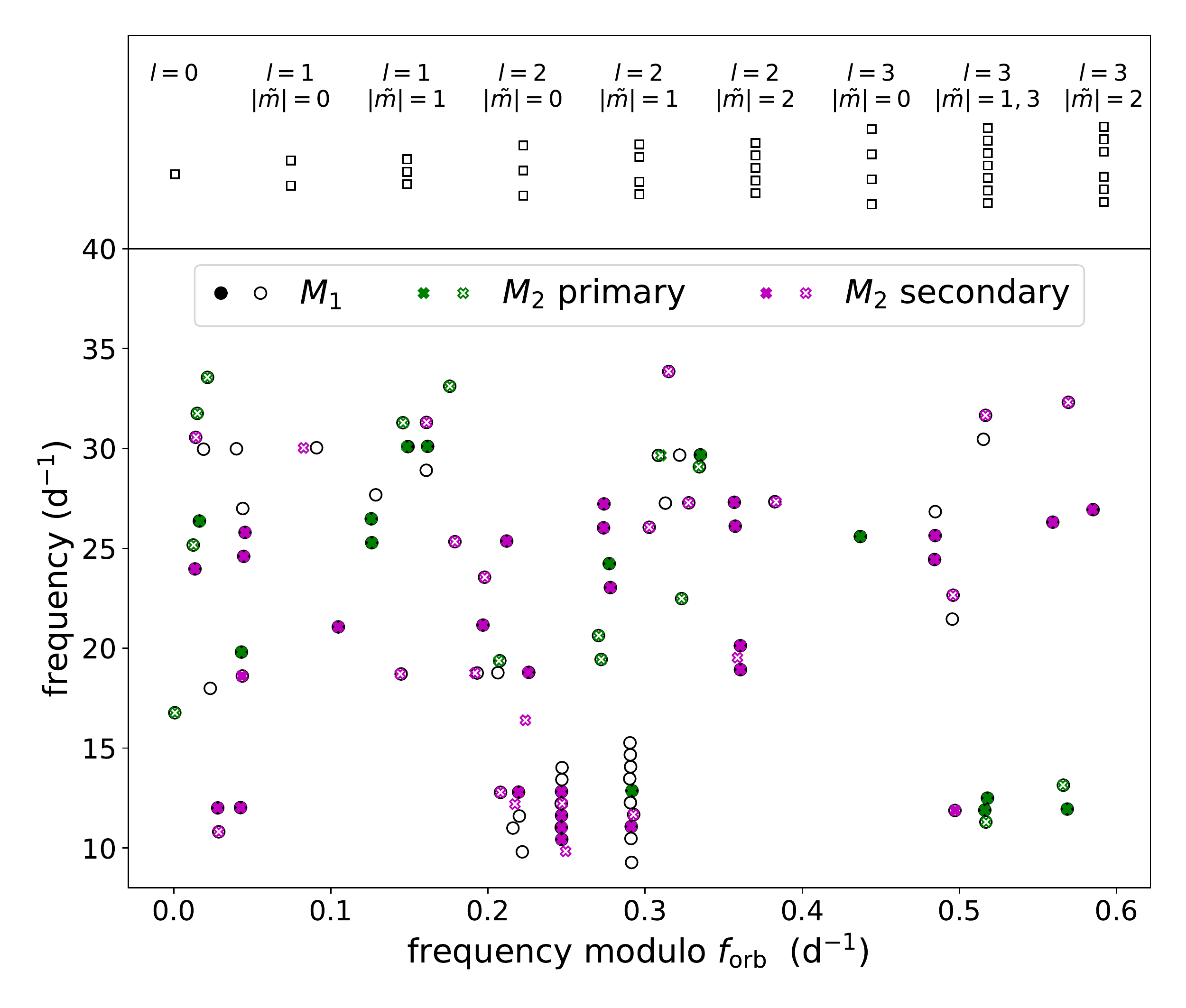}
      \caption{Amplitude spectrum modulo the orbital frequency to visualise the evidence for tidally perturbed modes. Top panel: The signature of multiplets expected from the theory of of perturbed pulsation corresponding to different values of $l$ and $|\tilde{m}|$. Multiplets are plotted according to the tables in \citet{smeyers2005} and \citet{balona2018}. Bottom panel: All significant frequencies from the traditional model and our toy model above 8 \cd modulo the orbital frequency. Frequencies from he traditional model are shown as black circles were filled circles indicate the frequencies with amplitudes above $165$~ppm and open circles frequencies below that amplitude level. Frequencies from our toy model are shown as coloured crosses: green crosses correspond to frequencies assigned to the primary component while magenta crosses correspond to frequencies assigned to the secondary component. Low amplitude modes are depicted as open crosses. 
              }
         \label{appFig:8:perturbed_modes}
\end{figure}

 Figure \ref{appFig:8:perturbed_modes} shows all significant frequencies from the traditional model and this new model above 8 \cd modulo the orbital frequencies. On multiple occasions, the multiplet in the latter is different. While for example F1 is part of multiplet of ten frequencies when modelling the light curve with traditional model, we only find a triplet in this approach. Furthermore, two low amplitude modes surrounding F2 are not anymore significant.
 
 Many more small amplitude frequencies are significant in the traditional model but are not significant in this different approach. Therefore, the signature of the multiplet is different, which partly leads to different mode identification.
 
 From this experiment, it is clear that amplitude modulation caused by changing light ratio in an eclipsing binary system plays a major role in the creation of frequency multiplets. While tidally perturbed modes should, in principle, allow mode identification, it is important to include a proper model for the amplitude variations.

\section{Tidally perturbed modes versus amplitude variations}
\label{App:pert_vs_var}
\begin{figure}
   \centering
   \includegraphics[width=\linewidth]{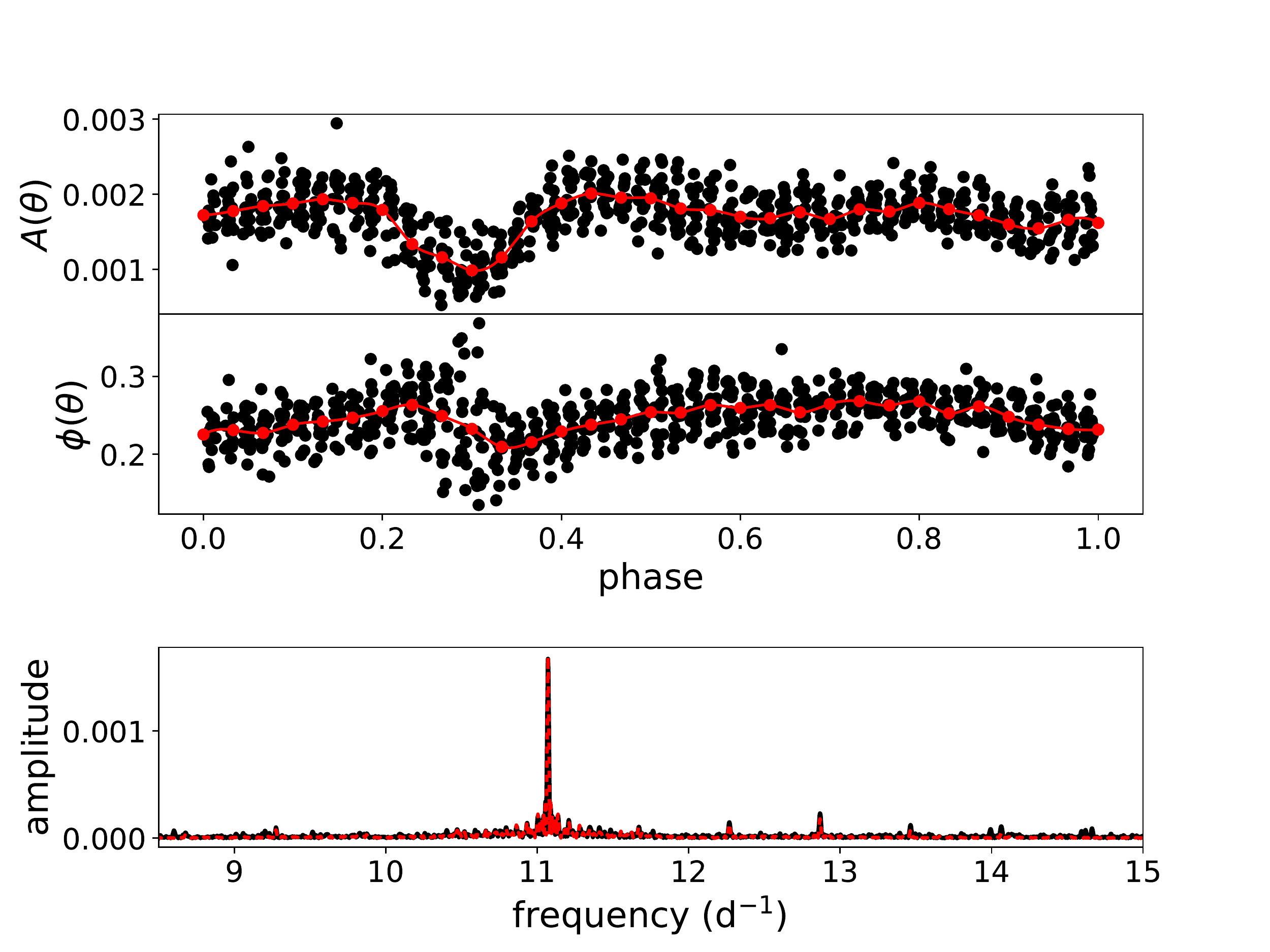}
      \caption{Amplitude and phase variations obtained when assuming the multiplet around F1 is solely due to an individual frequency at F1. Top panel: amplitude variations obtained from fitting F1 to subsets of the multiplet light curve (black points), the binned values (red points), and the cubic spline interpolation used as model (red line). Middle panel: same as the top panel but for phase variations. Bottom panel: Amplitude spectra of the multiplet light curve (black line) and of the resulting model flux (dashed red line). 
              }
         \label{F1_amp_variations}
\end{figure}

\begin{figure}
   \centering
   \includegraphics[width=\linewidth]{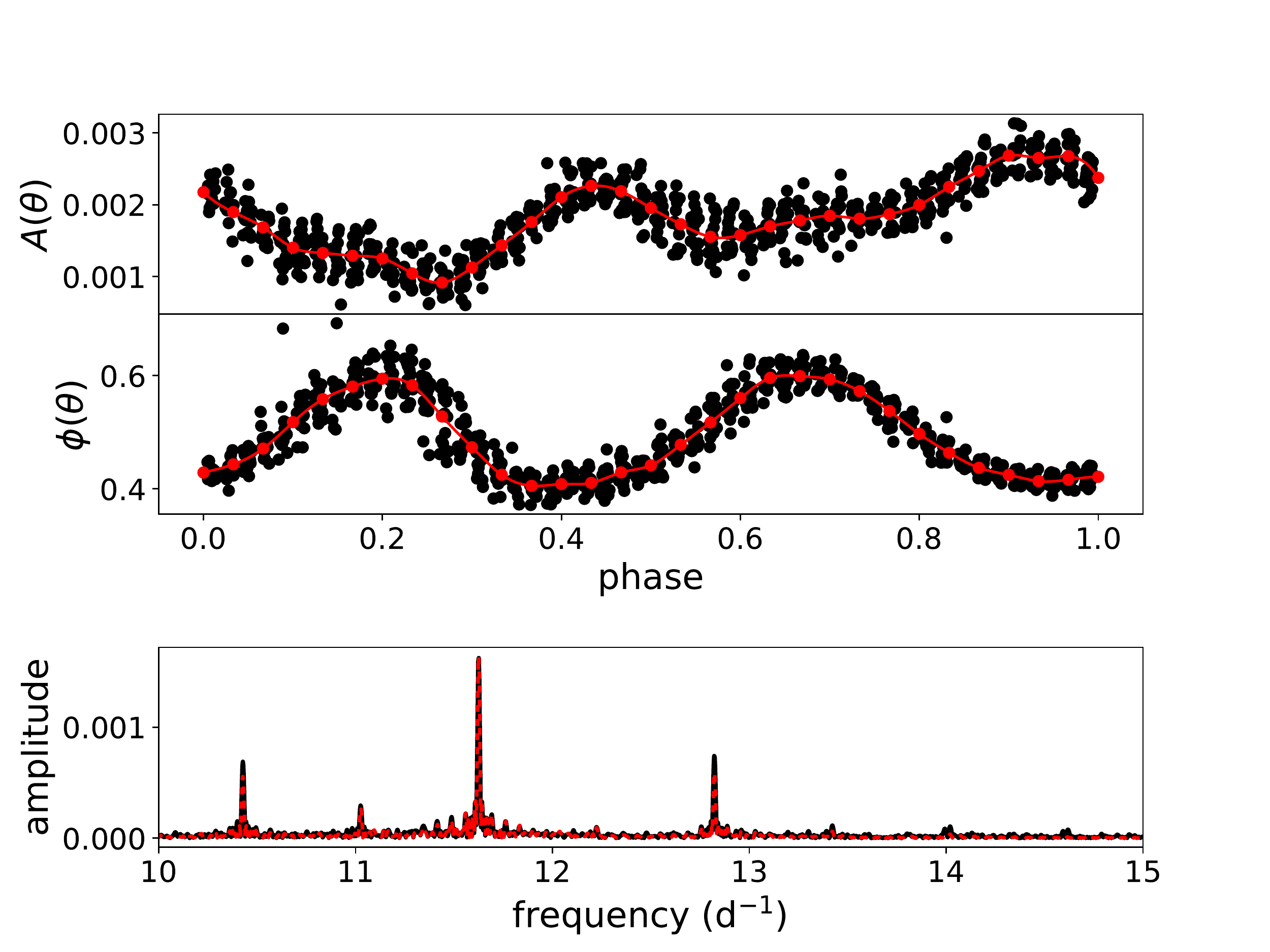}
      \caption{Same as Figure \ref{F1_amp_variations}, but for the multiplet around F2.
              }
         \label{F2_amp_variations}
\end{figure}

\begin{figure}
   \centering
   \includegraphics[width=\linewidth]{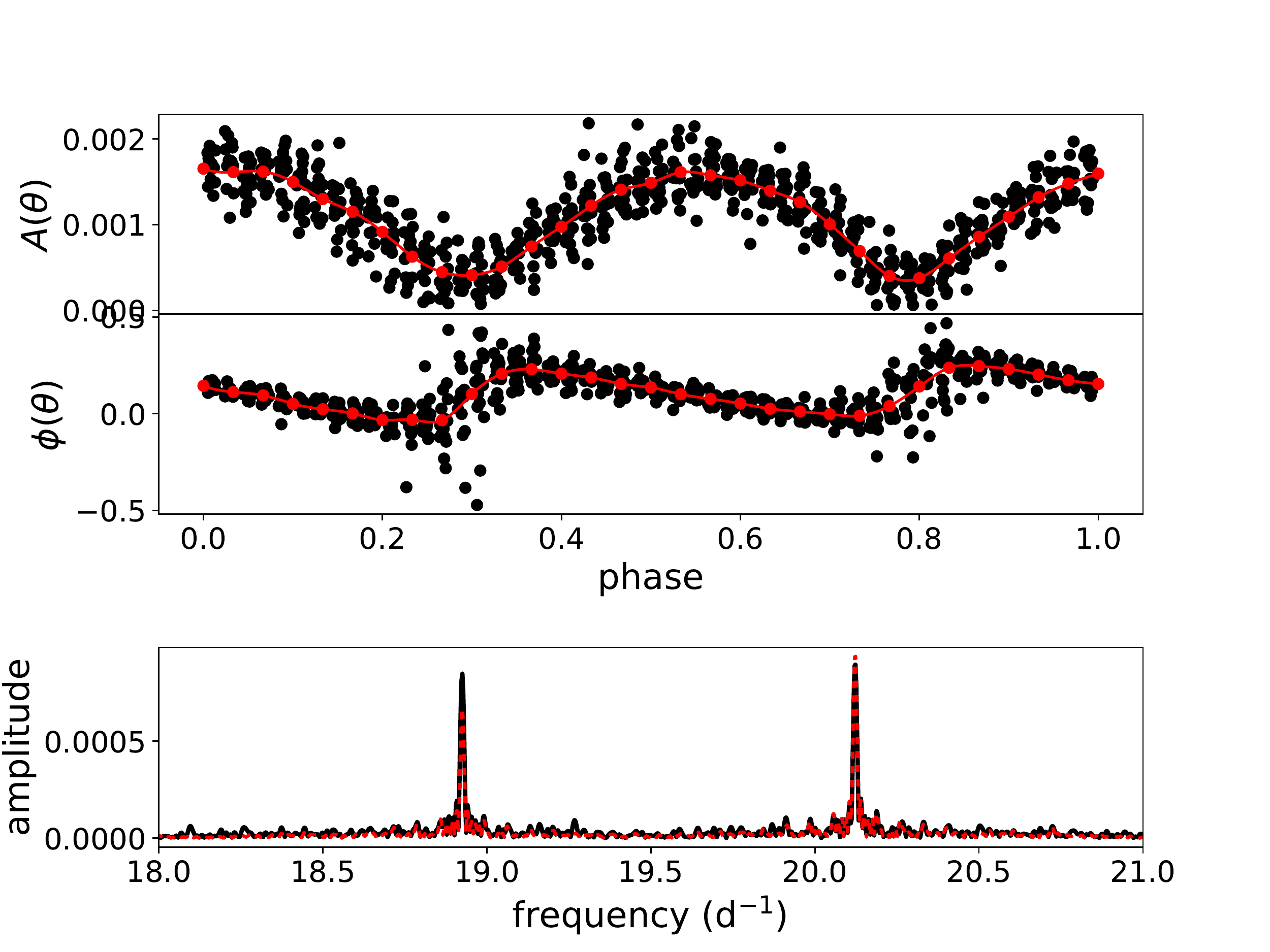}
      \caption{Same as Figure \ref{F1_amp_variations}, but for the doublet consisting of F4 and F5. 
              }
         \label{F4_F5_amp_variations}
\end{figure}

Figures \ref{F1_amp_variations} to \ref{F4_F5_amp_variations} show the result of our investigation concerning whether amplitude and phase variations can explain the amplitude spectrum of the multiplets (see Sect. \ref{pert_vs_var}).

\newpage

\end{document}